\documentclass[12pt]{article}
\usepackage{makecell}
\usepackage{booktabs}

\usepackage[flushleft]{threeparttable}
\usepackage{caption}
\usepackage{graphicx}
\usepackage{ragged2e}
\usepackage{lscape}
\usepackage{verbatim}
\usepackage{times}
\usepackage{floatrow}
\usepackage{adjustbox}
\usepackage{subcaption}
\newfloatcommand{capbtabbox}{table}[][\FBwidth]

\usepackage{blindtext}

\usepackage[title,toc,titletoc,page]{appendix}
\usepackage{threeparttable}
\usepackage{floatrow}
\usepackage[round]{natbib}
\usepackage{hyperref}

	\title{Power in the Pipeline\thanks{We wish to thank Chris Blattman, Ruben Durante, Björn Gehrmann, Gabriele Gratton, Bard Harstad, Matthew O.\ Jackson, Melika Liporace, Maria Petrova, Marina Petrova, Paolo Pin, Michael Porcellacchia, Fernando Vega Redondo, Camille Terrier, and seminar participants at Bocconi University, University of Padova, Monash University webinar, Oslo workshop on sanctions, University of Lausanne, the CEPR RPN workshop and the Kiel CEPR geopolitics and economics workshop for very useful comments. We gratefully acknowledge support from the European Research Council grants 694583 and 677595.
 The usual disclaimer applies. }}

\author
{ Quentin Gallea,$^{1,2} $  Massimo Morelli$^{3,4,5,6}$, \\ Dominic Rohner$^{1,2,6}$\\
\\
\normalsize{$^{1}$ Department of Economics, University of Lausanne. }\\
\normalsize{$^{2}$ Enterprise for Society (E4S) Center.}\\
\normalsize{$^{3}$ Department of Economics, Bocconi University.}\\
\normalsize{$^{4}$ IGIER.}\\
\normalsize{$^{5}$ Baffi CAREFIN.}\\
\normalsize{$^{6}$ CEPR.}\\
}

\date{\today}

\begin{document}


\baselineskip24pt


\maketitle

\begin{abstract}
This paper provides the first comprehensive empirical analysis of the role of natural gas for the domestic and international distribution of power. The crucial role of pipelines for the trade of natural gas determines a set of network effects that are absent for other natural resources such as oil and minerals. Gas rents are not limited to producers but also accrue to key players occupying central nodes in the gas network. Drawing on our new gas pipeline data, this paper shows that gas {\it betweenness-centrality} of a country increases substantially the ruler's grip on power as measured by leader turnover. A main mechanism at work is the reluctance of connected gas trade partners to impose sanctions, meaning that bad behavior of gas-central leaders is tolerated for longer before being sanctioned. Overall, this reinforces the notion that fossil fuels are not just poison for the environment but also for political pluralism and healthy regime turnover.

\bigskip

\noindent {\bf Keywords:} Natural gas network, betweenness-centrality, asymmetries, pipelines, regime durability, sanctions, democracy.

\noindent {\bf JEL codes: C33, D74, F51, Q34}

\end{abstract}


\section{Introduction}

Gas markets and their political consequences could not be more topical and timely in current geopolitics and international relations. Yet, the political consequences of the natural gas trade network have not been adequately studied by political economists. While there exists quite some empirical evidence on political side effects of oil extraction (see e.g. \citet{cotet2013oil,lei2014giant,caselli2015geography,morelli2015resource}), the political impact of natural gas abundance and trade is likely to be extremely different. A key characteristic of natural gas is that it is much cheaper to transport through pipelines than with any alternative technology. This gives key strategic power to central nodes in the pipeline network, whereas for oil it is easier to substitute suppliers and intermediaries.\footnote{While oil is primarily transported via tankers, most natural gas is transported through gas pipelines (\citet{petrol2010bp}, \citet{rodrigue2016geography}).} This is well illustrated by the discussion inside the European Union on sanctions against Russia following its invasion of Ukraine in Spring 2022: While rarely anyone is concerned about the costs for Europe of sanctioning Russian \emph{oil} (for which several substitutes exist), boycotting Russian \emph{gas} is generally considered much more painful for European economies, given the great difficulties to substitute it in the short-run.

This paper aims to study whether the particular dependence or asymmetry implicit in gas trade has determined greater survival in office of leaders within countries that have been central in the network. There are various reasons to think that in countries occupying a central node in the network of international gas pipelines it is easier for a given regime to cling to power. First of all, key players on the gas market may be able to escape some types of international sanctions, and lower international scrutiny may fuel regime survival. Second, being a key intermediary in the world gas trade is lucrative and may allow politicians to corrupt or weaken oppositions and hollow out institutions. In this paper we investigate these conjectures empirically. To the best of our knowledge, our paper contains the first statistical analysis of the impact of arguably exogenous changes in gas centrality on regime survival. We draw on our novel data set on gas pipeline locations, covering Africa, Asia, Europe and Oceania with 265 different pipelines spread out over 67 countries. As far as we know, our dataset contains every international natural gas pipeline connection built over those four continents from 1963 (first international connection recorded) to 2018. A crucial feature of this novel dataset is that we have also collected information on construction years, which other datasets typically lack (see detailed discussion of data collection below). Controlling for batteries of fixed effects and for local degree centrality, we exploit far away changes to the overall network structure that lead to exogenous changes in the (betweenness) centrality of a given node. For example, with the construction of the European southern corridor (pipeline network from Azerbaijan to Italy), the centrality of Belarus and Ukraine has changed. 

Our main result is that an increase in gas centrality of a country consolidates internal power, manifested by a higher leader survival probability. Our overall statistical findings for a panel data set with a large number of countries are well illustrated by the example of gas-central Belarus, where President Lukashenko has been in office since 1994, making him the longest-sitting European president. In our econometric analysis we show that this case is not an exception, but rather symptomatic of  a general pattern present for a large number of countries and years.

Importantly, we find evidence for a key mechanism behind power consolidation: increases in gas centrality have led on average to a lower frequency of sanctions against such more central countries (in particular sanctions to restrict arms trade and to prevent war). This is well illustrated for example by the fact that after the Russia-Ukraine crisis in 2014 it took a full-blown large scale invasion in 2022 by Russian forces to make European countries (who are large consumers of Russian gas) willing to step up the scale of the sanctions. The comprehensive statistical analysis demonstrates that this example is the rule and not an exception, and allows us to assess also the role of other potential mechanisms, which we find to not play a crucial role.

In terms of related literature, first and foremost the work on various facets of the so-called "natural resource curse" is relevant (see the survey of \citet{van2011natural}).
Fossil fuels, and in particular oil abundance, have been linked to various political ills such as corruption and mismanagement (\citet{caselli2013oil}), civil conflicts (\citet{lei2014giant,morelli2015resource}), mass killings (\citet{esteban2015strategic}), military spending and aggressive behavior in petro-states (\citet{cotet2013oil,hendrix2017oil}), and militarized interstate disputes and wars (\citet{acemoglu2012dynamic,colgan2013fueling,koubi2014natural, caselli2015geography}). The nexus between democracy and natural resources is less clear-cut, with various articles finding that resource abundance threatens democracy (\citet{ross2001does,andersen2014big}), while some work finds countervailing results (\citet{haber2011natural}).

A second related  literature is  the economics of networks, where military alliance and enmity networks have been linked to civil and interstate conflicts (\citet{dziubinski2016conflict,konig2017networks,fearon2018arms,gallea2019arming}).
Positions in geographical and trade networks have been related to conflict (\citet{polachek1980conflict, martin2008make, rohner2013war,gallea2021globalization,mueller2022ethnic}) but not to internal and external power dynamics within and across countries.

The third and most related literature to our paper is the one linking natural resources and regime durability. As far as the link between oil and regime durability is concerned, an increased regime longevity for oil producers has been documented for 26 African states (\citet{omgba2009duration}) and for a sample of 106 dictators (\citet{crespo2011oil}). Also \citet{brausmann2020resource} show that large hydrocarbon discoveries lower the loss of power hazard faced by an autocrat.  Several studies point out that the relationship is not clear-cut, and depends very much on the type of natural resource in question: \citet{de2010leader} find contrasting effects in countries with autocratic versus democratic leaders. Similarly, \citet{andersen2013oil} find that wealth derived from natural resources affects political survival in intermediate and autocratic, but not in democratic, countries.
While oil and non-lootable diamonds are associated with positive effects on the duration in political office, minerals are associated with negative duration effects. Further, as shown by \citet{nordvik2019does}, oil price shocks fuel coups in onshore-intensive oil countries, while deterring them in offshore-intensive oil countries.\footnote{For a recent survey covering outside interventions and sanctions, see \citet{rohner2022mediation}.}

Our paper contributes to this existing literature in several respects. First, while existing work overall focuses on oil, we study natural gas which has received much less attention. Second, the key importance of gas transport through pipelines allows us to study a new type of exogenous shock  -- namely sharp changes to betweenness centrality of intermediary countries (arising due to the addition of far away nodes, when controlling for local degree centrality). Third, the key salience of international gas trade networks allows for a subtle investigation of mechanisms at work.

The remainder of the paper is structured as follows. The next Section \ref{Sec_methods} presents the context, data and methods. In particular, Subsection \ref{Sec_context} provides a short overview of the basics of gas trade and interstate disputes.
In Subsection \ref{Sec_data} we describe the data used, with a special focus on our novel data set on gas pipelines.
Next,  Subsection \ref{Sec_empirics} outlines the identification strategy and presents the specification that we run.
Section \ref{Sec_results} is dedicated to displaying our main results. Finally, Section \ref{Sec_conclusion} concludes. Supplementary material is provided in several Appendices.

\section{Method}\label{Sec_methods}

\subsection{Gas trade context}\label{Sec_context}

In order to motivate the empirical approach followed, we shall start by briefly discussing the particularities of the gas trade and transport. While the trade of oil mostly takes place using tanker ships crossing the oceans, gas trade in contrast relies predominantly on a network of gas pipelines. Indeed 78\% of the natural gas trade in 2008 was done by pipelines (\citet{petrol2010bp}) while 38\% of crude oil is transported by pipelines, trucks or trains in 2015 (\citet{rodrigue2016geography}). The transport cost differential is a key reason for this difference: transporting gas via pipelines is much cheaper.

This is due to the very large costs and complex infrastructure required to liquifying and re-gasifying natural gas. Liquefied Natural Gas (LNG) has an advantage for large quantities (more than one million cubic meter) shipped
for long open sea distance (\citet{rodrigue2016geography}), with an indifference threshold often argued to be between 4,000 and  6,000 km (where the exact threshold of course depends on prices and technology).\footnote{Energy Brain Blog, last accessed 3rd of May 2020: \url{https://blog.energybrainpool.com/en/tutorial-gas-market-6-natural-gas-transportation-and-storage/}.}
This makes   pipelines the only economically viable way of transporting natural gas for most countries. Hence, a central node in the gas network can have large bargaining power, obtaining a high mark-up even if it doesn't have gas endowments.
Thus, intermediaries may grab considerable rents.

Further information on the gas and oil trade and price formation is relegated to the Appendix \ref{context_gas}.

\subsection{Data and descriptive statistics}\label{Sec_data}

\subsubsection{Overview on general structure}

We have built a country-year dataset from 1993 to 2018 covering Africa, Asia and Europe, composed of 168 countries. The international pipeline network began to grow during the 90s, creating a giant multi-continent network, and many countries founded after the dissolution of the USSR are of particular interest (Central Asia, Eastern Europe and the Caucasus). The dataset starts in 1993, the year when the current constitution of Russia was implemented.
The end date of our dataset varies from 2008 to  2018, depending on data availability of the outcome variables. We have restricted our analysis to Africa, Asia and Europe, as it formed a giant connected network already back in the early 1990s.\footnote{The continents are all geographically interconnected, and hence belong to the same network, while the remaining continents form separate networks.}

\subsubsection{Natural gas international pipelines}

\begin{figure}[ht!]
     \centering
       \caption{Natural gas pipeline international interconnections in 2018}\label{fig:map}
         \includegraphics[width=\textwidth]{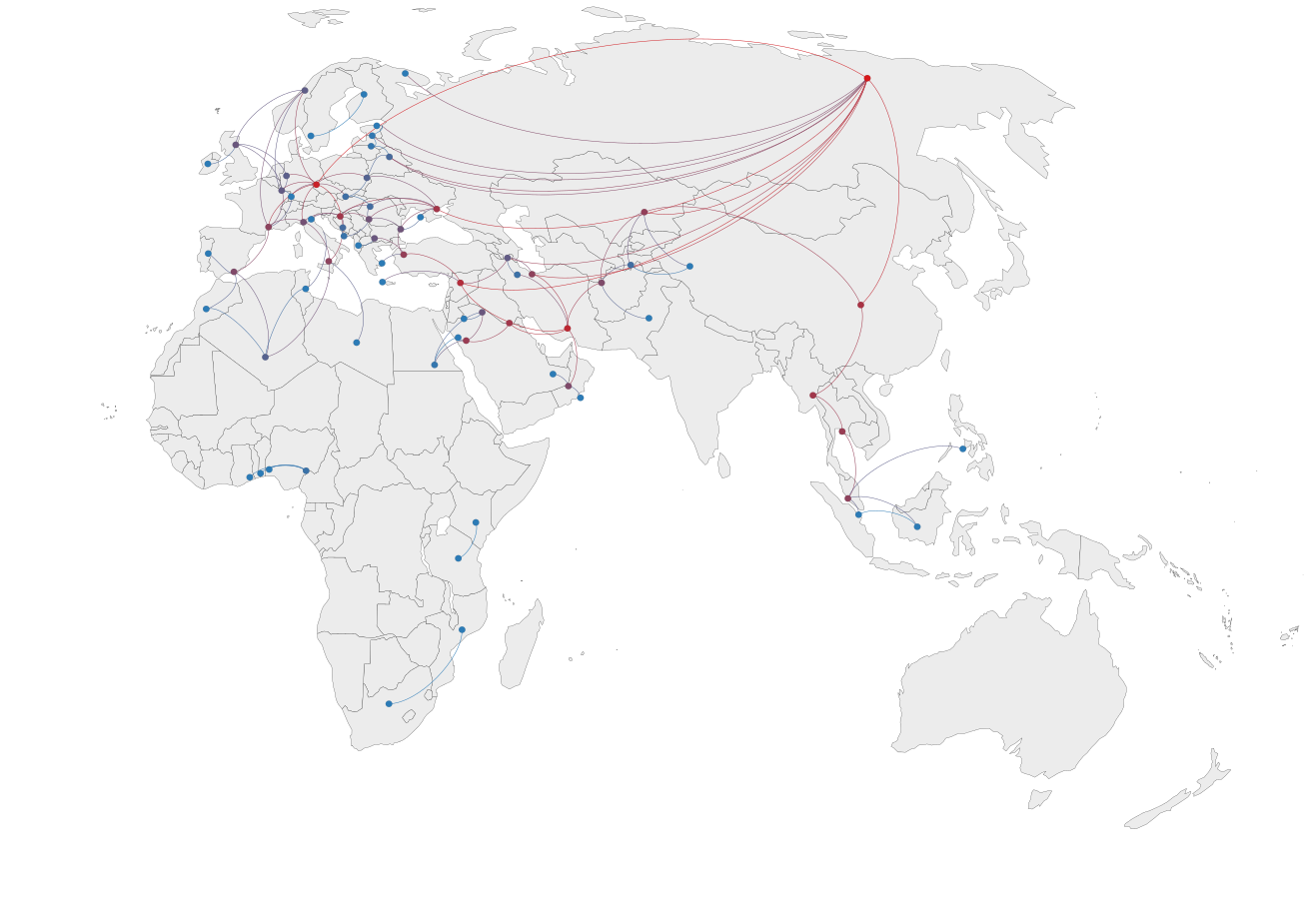}
                 \footnotesize \emph{Note:} The color of the nodes represents the betweenness centrality (from blue for low values to red for high values).
\end{figure}

We have collected the year of the first natural gas connection between countries from 1962 to 2018. The dataset contains a total of 311 transnational projects among which 256 are delivering gas in 2018 (see Figure \ref{fig:map}). We first geo-coded the location of natural gas pipelines using the Energy Web Atlas. Then, heavily relying on the geo-coded data set, we searched for the year where the pipelines were commissioned, using several resources: website from the operator or shareholder, journals focused on energy (Pipeline and Gas Journal, Hydrocarbons-technology.com) and others. Many of these pieces of information are hard to access, due to their confidential, geo-strategic nature.

A unique feature of our digital network data is that we have unique time-varying information on the network structure. This represents a large value added vis-à-vis most existing data that focus on one point in time. The importance of taking into account the time dimension is key, as between 1993 and 2010 there was roughly a doubling of connections within- and between-continents. This is shown in Appendix \ref{App_desc_stat}, which displays descriptive summary statistics on pipeline connections. In particular, Table \ref{tab:pipeline_distribution} summarizes the number of connections recorded in our dataset by continent for 1993 and 2010.

\subsubsection{Dependent variables}

First of all, we rely on two widely used datasets to measure the durability of the leader: Democracy and Dictatorship Revisited \citep{cheibub2010democracy}, and Archigos \citep{goemans2009introducing}. We have generated dummies taking the value one if the head of the state changed in a year $t$ in country $i$ (i.e. leader turnover). Further, we draw on the Global Sanctions Data Base for information on international sanctions \citep{kirilakha2021global}. Moreover, we use democracy scores from three datasets: The Polity V dataset \citep{polity5}, the V-Dem dataset \citep{pemstein2018v}, and the Freedom House dataset \citep{house2022freedom}. In all three cases (greater) democracy is reflected by higher scores.

Next, we use World Bank data for the GDP per capita in purchasing power parity in constant 2017 USD \citep{bank2020world}. Finally, to explore the quality of the government we use two datasets, namely \cite{prs2021}, as well as \cite{hanson2021leviathan}. In \cite{prs2021}, the quality of government is a scaled ($\in[0-1]$) index based on Corruption, Law and Order, and Bureaucracy Quality (higher values corresponding to higher quality). Related to this, the state capacity index from \cite{hanson2021leviathan} aggregates information on extractive capacity, coercive capacity, and administrative capacity.

\subsubsection{Control variables}

In addition to country and year fixed effects, we include in certain specifications four control variables:
the log of GDP per capita in purchasing power parity constant 2017 USD, the log of population, the log of natural gas rents as share of GDP as well as the log of oil rents as share of GDP (all from \citep{bank2020world}).\footnote{The four controls are log transformed due to their highly skewed right distribution.} In the Appendix we also provide a sensitivity analysis, controlling for variants of internal conflict (data from UCDP \citep{gleditsch2002armed}).

\subsection{Empirical Strategy}\label{Sec_empirics}

Our empirical model aims to predict various political and economic outcomes using network centrality measures.

\begin{eqnarray}\label{eq:baseline}
Y_{it} =  \beta_1 ln(BC_{i(t-1)})+ \beta_2 DC_{i(t-1)} + X' \lambda + FE_i + FE_t + u_{it}
\end{eqnarray}
with $i$ and $t$ standing respectively for country and year, BC for betweenness centrality, and DC for degree centrality. $X$ is a vector of controls containing: the log of GDP per capita in purchasing power parity constant 2017 USD, the log of population, the log of natural gas rents as share of GDP, as well as the log of oil rents as share of GDP. $FE_i$ and $FE_t$ stand for country and year fixed effects. $u_{it}$ is an error term clustered at the country level.

Betweenness Centrality (BC) of a country counts the shortest paths from producers to final consumers that pass through that country. Formally, we define a node $i$'s BC as the proportion of shortest paths between any other two nodes that pass through node $i$.

\begin{equation}
\label{BC1}
BC(v)=\sum_{s\neq v\neq t} \frac{\sigma_{st}(v)}{ \sigma_{st}}
\end{equation}
where $ \sigma _{st}$ is the total number of shortest paths from node $s$  to node $t$ and $ \sigma _{st}(v)$ is the number of those paths that pass through vertex $v$.

To ensure that the measure of betweenness centrality is comparable between unconnected networks of different sizes (e.g. European vs. South African networks), we use a standard approach to normalize this measure. Formally, the normalization is defined as follows:
\begin{equation}
BC^n=\frac{2 BC}{(n-1)(n-2)}
\end{equation}

where $BC^n\in\{0,1\}$ is the normalized $BC$, $BC$ is the unnormalized betweenness centrality and $n$ is the number of vertices in the network.

Given that the distribution of BC is skewed, we use the natural logarithm (ln). As there are country-years with a BC of zero, for which the ln would be non-defined, we apply the following formula: ln(BC+minium value of BC in the sample). This slight rescaling guarantees that the ln is always well defined.

Degree Centrality (DC) of a country instead counts the number of direct connections of the country in the gas network.

The political and economic situation of country $i$ (captured by some of the outcome variables we consider) might affect the decision to build a new pipeline and hence influence our centrality measures. Hence, interpreting the coefficient of DC in a causal manner would be dangerous, as it could be driven by endogenous confounders. The inference strategy we adopt is thus to focus on changes in BC while controlling for DC (similarly to \cite{donaldson2016railroads}). This amounts to
filtering out events taking place in or near the country in question and exploit solely events far away. Intuitively, one can think of BC as a {\it global} measure of centrality of a country, whereas DC is a {\it local} measure, since it counts only the direct links. Therefore having both BC and DC in the regression model allows to study the net effects of global centrality (BC), net of the local environment (DC), and the local confounders associated to it.

\section{Results}\label{Sec_results}

The key findings will be summarized in four figures, whereas all underlying regression tables and further results and robustness checks will be relegated to the Appendices \ref{app:tables} and \ref{app:robustness}.

\begin{figure}[ht!]
     \centering
         \includegraphics[width=\textwidth]{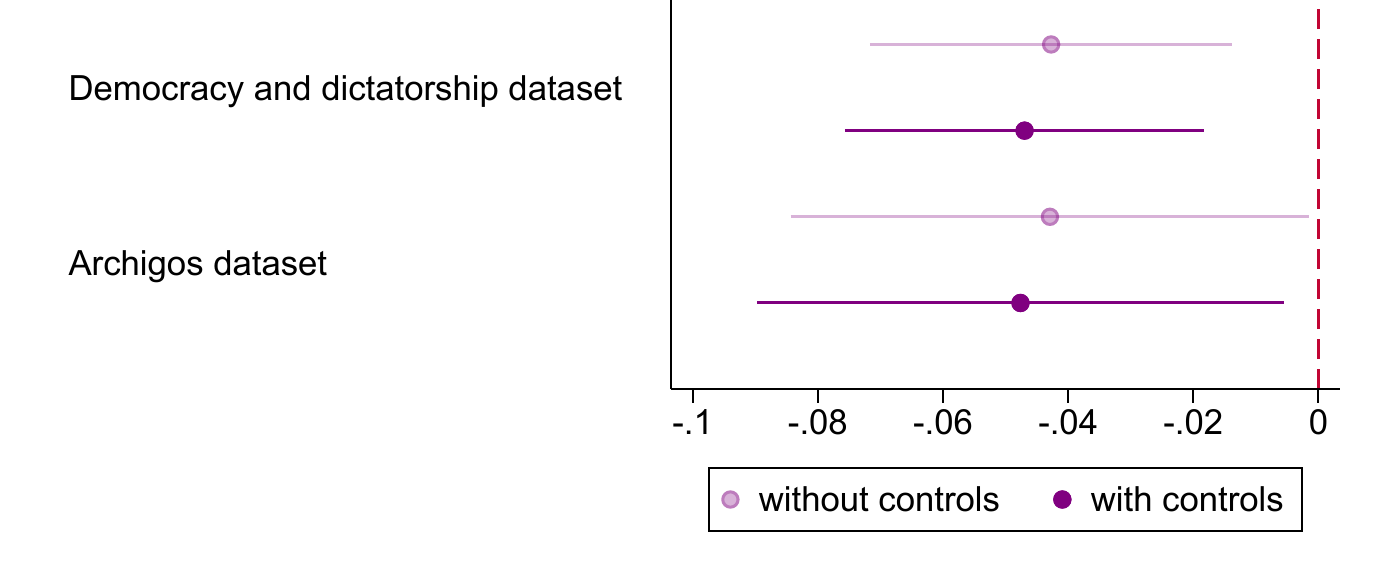}
         \caption{Leader turnover}\label{fig:main}
		\justifying
         \footnotesize  \emph{Note: }Value and 95\% confidence intervals for the betweenness centrality coefficient. Linear Probability Models estimated with OLS. Standard error clustered at the country level. The unit of observation is country-year and the dataset spans from 1993 to 2018 covering Africa, Asia and Europe. The outcome is a dummy taking the value one if the executive leader changed during the year. Betweenness centrality corresponds to the lagged natural logarithm of the natural gas pipeline betweenness centrality for country $i$. The vector of controls includes: degree centrality for country $i$, the log of the population, the log of the GDP per capita in ppp, the log of natural gas rents as well as the log of oil rents both in percent of the GDP.
\end{figure}

Figure \ref{fig:main} displays the main results. We find that the likelihood of leader turnover in office is significantly reduced when a country is gas central. This holds across both different datasets for leader survival in office, i.e. "Democracy and dictatorship dataset" and Archigos, and for both specifications without controls and once a battery of control variables are included. Reassuringly, the inclusion of controls --if anything-- strengthens the results. Quantatively, the baseline result means that if ln(BC) increases by 1 then government turnover gets reduced by 0.05 (i.e. 5 percentage points), which corresponds to a quarter of the baseline risk of 0.2.

\begin{figure}[ht!]
     \centering
         \includegraphics[width=\textwidth]{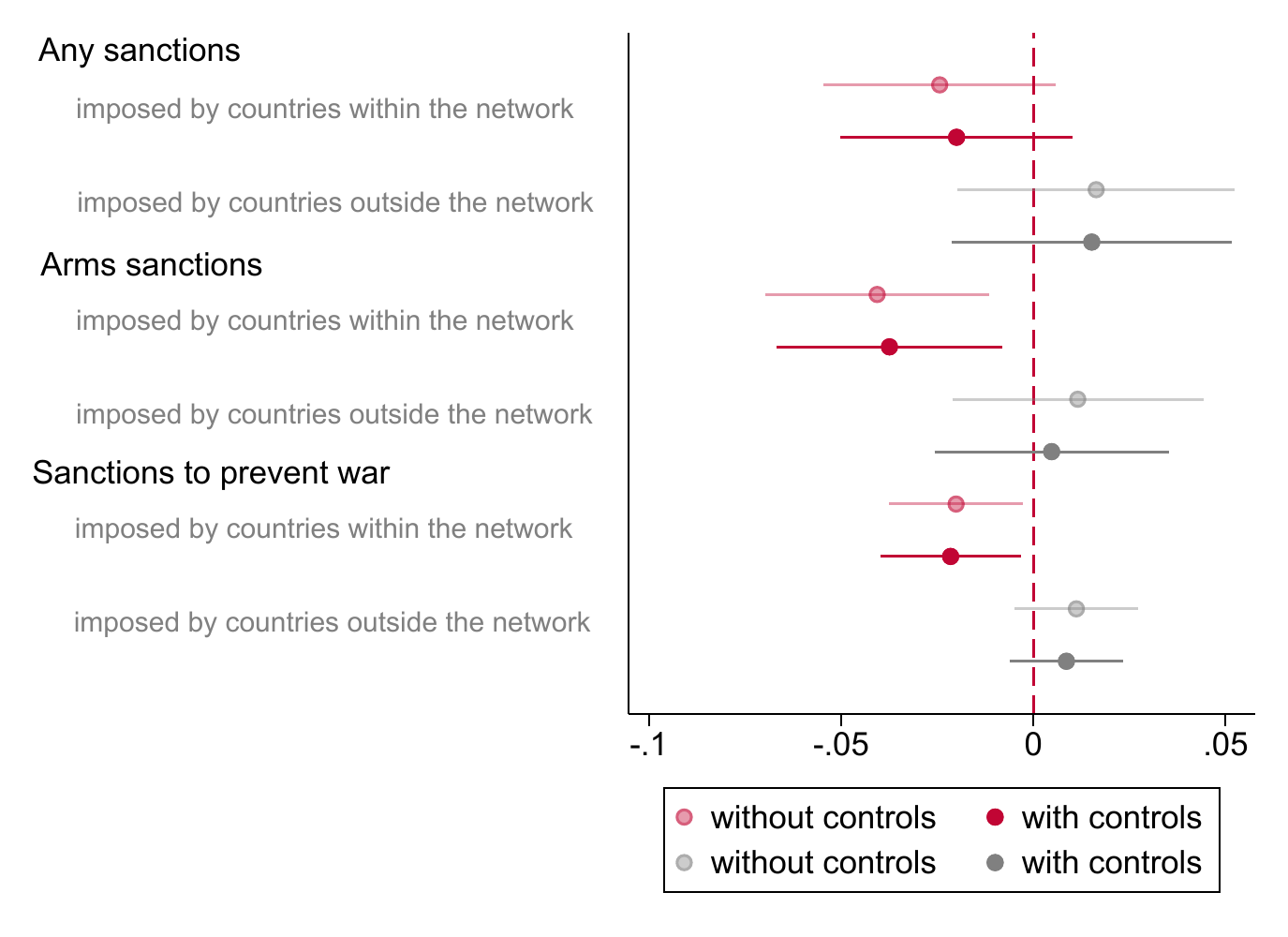}
         \caption{Sanctions}\label{fig:sanctions}
	\justifying
         \footnotesize \emph{Note: }Value and 95\% confidence intervals for the betweenness centrality coefficient. Linear Probability Models estimated with OLS. Standard error clustered at the country level. The unit of observation is country-year and the dataset spans from 1993 to 2018 covering Africa, Asia and Europe. The outcome is a dummy taking the value one if country $i$ was subject to sanctions in year $t$. There are two dimensions with respect to the outcome. First, we use three different types of sanctions: any type of sanctions; then arms sanctions; and finally sanctions to prevent war. Second, we restrict attention to different sets of countries imposing the sanctions: sanctions imposed by countries connected to country $i$ in the natural gas pipeline network and sanctions imposed by countries not connected to country $i$ within the pipeline network. The vector of controls includes: degree centrality for country $i$, the log of the population, the log of the GDP per capita in ppp, the log of natural gas rents, as well as the log of oil rents both in percent of the GDP.
\end{figure}

Figure \ref{fig:sanctions} investigates whether the increased geo-political importance makes on average a gas-central country less (or later) subject to sanctions, vis-a-vis countries who are connected to the same gas network. Overall, for any sanctions we find the expected negative coefficient, but statistical significance is narrowly missed. However, when zooming in on particular sanction types, we find that military-related sanctions (i.e. "arms sanctions" and "sanctions to prevent war") become, as expected, statistically significantly less likely to be put in place by network-connected countries against gas-central nations.\footnote{By "network-connected" we understand "being directly or indirectly connected". So if, e.g., Switzerland is connected to Austria and Austria is connected to Hungary, then the dyad Switzerland-Hungary counts as "network-connected". Note that the share of network-connected pairs increases over time, as the network has become increasingly dense and integrated over time (see Appendix Table \ref{tab:pipeline_distribution}).} This finding both holds for the "parsimonious" specifications without controls, as well as for the specification with the full battery of controls. Again, reassuringly, the coefficient is very stable to the inclusion of controls. Further, in the spirit of a placebo test, the probability of sanctions being imposed by countries outside the network is, if anything, increased but the coefficient is not statistically significant. Taken together, it turns out that indeed a range of sanctions are less likely to occur for gas-central countries and this is entirely driven by ``sanction hesitancy'' by countries within the gas network.

The next Figure \ref{fig:demo} investigates to what extent a reduction in democracy scores could be a potential channel of gas-centrality fostering regime durability. While the coefficients for all three datasets considered (Polity IV, V-Dem, Freedom House) have always the expected negative sign, they lack statistical significance.

\begin{figure}[ht!]
     \centering
         \includegraphics[width=\textwidth]{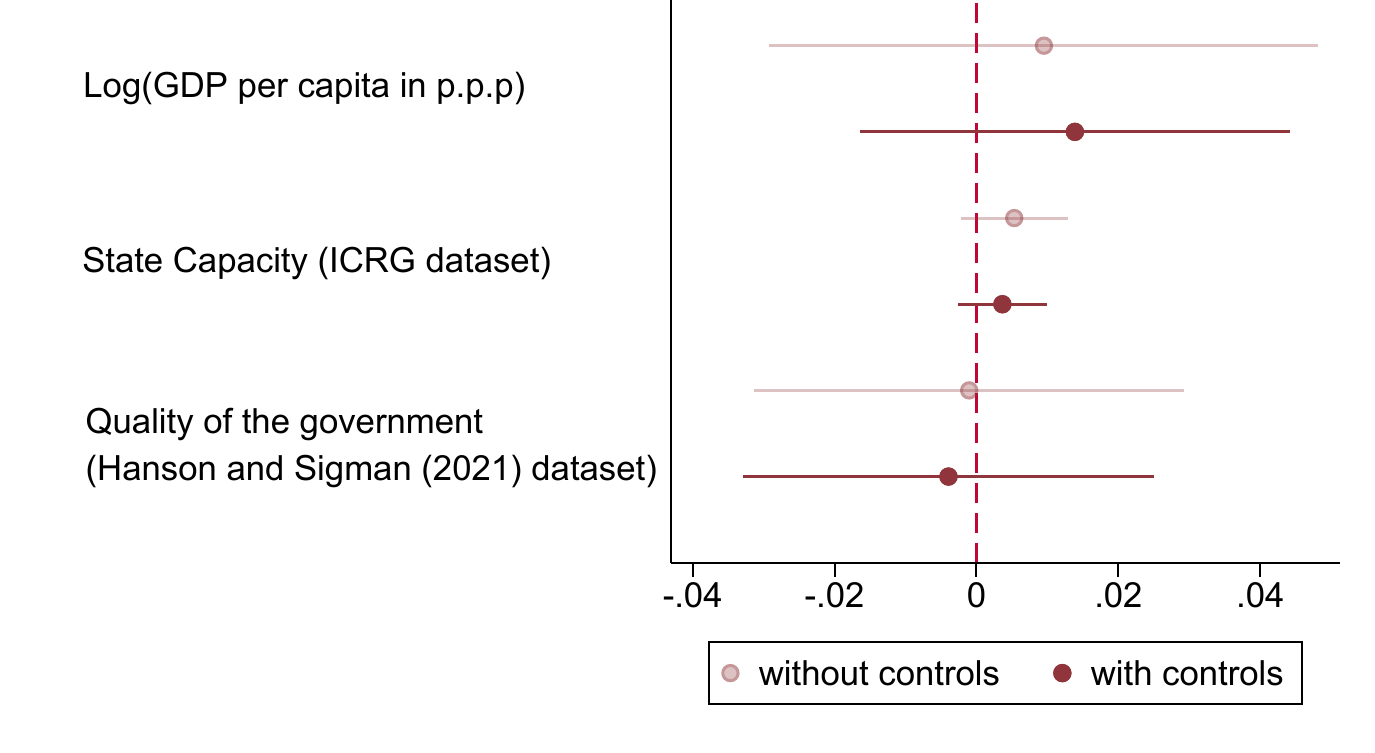}
         \caption{Other outcomes: Value and 95\% confidence intervals for the betweenness centrality coefficient}\label{fig:gdp}
         \justifying
         \footnotesize  \emph{Note: }Value and 95\% confidence intervals for the betweenness centrality coefficient. Linear Probability Models estimated with OLS. Standard error clustered at the country level. The unit of observation is country-year and the dataset spans from 1993 to 2018 covering Africa, Asia and Europe. The outcome variables are log of GDP, state capacity and quality of the government. Betweenness centrality corresponds to the lagged natural logarithm of the natural gas pipeline betweenness centrality for country $i$. The vector of controls includes: degree centrality for country $i$, the log of the population, the log of the GDP per capita in ppp, the log of natural gas rents as well as the log of oil rents both in percent of the GDP.
\end{figure}

Then, in the last Figure \ref{fig:gdp} we ask the natural follow-up question on whether this longer regime survival may be driven by economic or other benefits yielded by gas centrality. Put differently, we investigate whether leaders can cling to power longer because gas actually helps them to do better. We find that this is not the case. For none of the outcome variables (GDP, state capacity, quality of government) we find any beneficial effect of gas centrality.

As mentioned above, all the results from Figure \ref{fig:main} to \ref{fig:gdp} are highly robust to the inclusion of a set of controls: the log of GDP per capita in purchasing power parity constant 2017 USD, the log of population, the log of natural gas rents as share of GDP as well as the log of oil rents as share of GDP. In particular, controlling for natural gas and oil rents highlights the relevance of our identification strategy. The coefficients are not statistically significant for the rents and the results are indeed driven by the network structure captured by our centrality measure. This highlights again the fundamental difference between gas (where network location is key) versus oil (where producer rents are most crucial). Appendix \ref{app:tables} displays in table form all regression results underlying Figures \ref{fig:main} to \ref{fig:gdp}. In the Appendix \ref{app:robustness} several sensitivity / robustness tests are performed.

\begin{figure}[ht!]
     \centering
         \includegraphics[width=\textwidth]{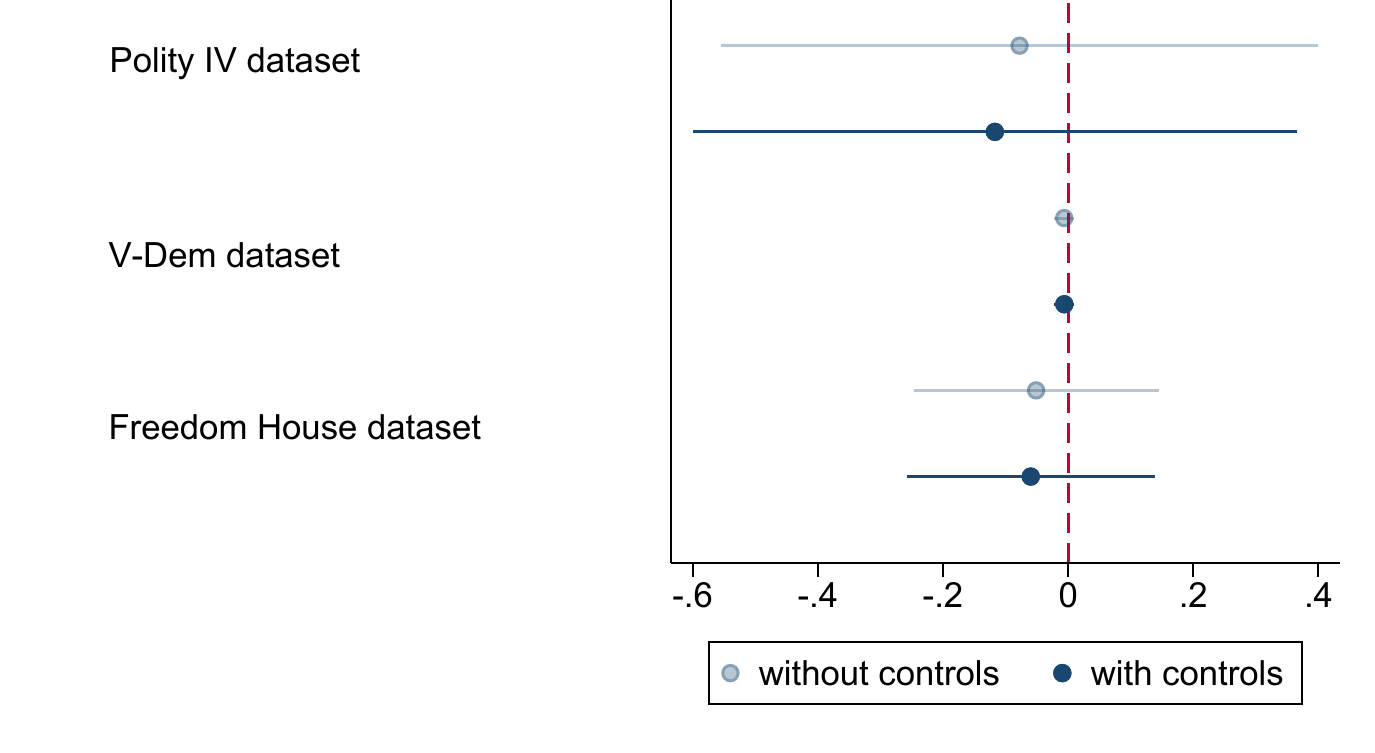}
                  \caption{Democracy: Value and 95\% confidence intervals for the betweenness centrality coefficient}\label{fig:demo}
\justifying
         \footnotesize  \emph{Note: }Value and 95\% confidence intervals for the betweenness centrality coefficient. Linear Probability Models estimated with OLS. Standard error clustered at the country level. The unit of observation is country-year and the dataset spans from 1993 to 2018 covering Africa, Asia and Europe. The outcome variables are democracy scores from, respectively, Polity IV, V-Dem and Freedom House (higher scores correspond to (greater) democracy). Betweenness centrality corresponds to the lagged natural logarithm of the natural gas pipeline betweenness centrality for country $i$. The vector of controls includes: degree centrality for country $i$, the log of the population, the log of the GDP per capita in ppp, the log of natural gas rents as well as the log of oil rents both in percent of the GDP.
\end{figure}

\clearpage

\section{Discussion}\label{Sec_conclusion}

This paper provides new results and insights on the resource curse and, in particular, on the special role played by natural gas in recent history. Increases in the centrality of a country in the gas trade network result in a higher probability of regime survival, and such a leader durability effect does not appear to be due to greater economic opportunities. Rather, we find that there is clear evidence that increased centrality of a country reduces the willingness of other countries in the same network to sanction the gas-central country. Put differently, gas centrality does not seem to increase the total welfare of the country but rather has a significant effect on the consolidation of power by the leaders.

Could one read Ukraine's political evolution through the lens of these results? The country used to have a high betweenness centrality in the gas market, which however was reduced by the creation of the Georgia-Turkey pipeline in 2007 and the strengthening of the direct pipeline potential between Russia and Germany. According to our results from the global statistical investigation, the impact of this relative loss of centrality would be higher government turnover and the country being less shielded from sanctions or other actions by countries in the same gas network. On another note, our findings also illustrate how hesitant/reluctant European countries have been to sanction gas imports from gas-central Russia after its invasion of Ukraine in Spring 2022.

Overall, one policy implication of our study is that if we want to live in a world where all countries are equally accountable for their actions and none are shielded from sanctions by gas centrality, we need to move away from fossil fuels and speed up the green energy transition \citep{garicano2022global}. This will allow to tackle two grand challenges at the same time: toxic politics and global warming.


\newpage

{\footnotesize
\bibliographystyle{apa}
\bibliography{bibli}
}

\newpage

\begin{appendices}

In what follows, several Appendices will provide additional explanations, results and robustness checks.

\section{Gas trade context -- additional information}\label{context_gas}

The current Appendix provides further information on the gas trade context. In the main text we discussed transportation for gas. Below we start by discussing the predominant means of transport for oil, which are very different. Even if oil travels through pipelines for part of the trades as well, holding a central position in the network is not so valuable, because if centrally located countries were to try to charge too high passage mark-ups, the sellers would switch to a tanker transportation as a close and relatively cheap alternative (as tanker shipping for oil is significantly cheaper than pipeline transportation for equivalent distance).\footnote{Energy Brain Blog, last accessed 3rd of May 2020: \url{https://blog.energybrainpool.com/en/tutorial-gas-market-6-natural-gas-transportation-and-storage/}.}
The above mentioned differences imply that  the really valuable positions in the oil network are the production nodes, whereas for the gas network it is possible that a central node without production or endowment could be even more attractive than the production nodes (especially if viable transport alternatives do not exist).

The 2019 International Gas Union Wholesale Gas Price Survey (\citet{IGU2019}) shed light on the gas price formation mechanism for pipeline transfers. In this annual global survey, the price is decomposed in several categories to pin down the percentage of the final price attributed to each category.
Part of the price is related to indices such as Henry-Hub, NBP or NYMEX; part of the price relates to the price of oil; and, most important for us, part of the price is due to bilateral monopoly relations, where large buyers and sellers set the price for a deal for a year or more. Not surprisingly, the part of the price due to bilateral monopoly is the most important in the Middle east ($\approx 80\%$), Former Soviet Union ($\approx 60\%$) and Africa ($\approx 40\%$), places where central actors are found. Since 2015 the distribution remained relatively stable.

Note in contrast that $0\%$ of the price formation of liquified natural gas (the type of gas traded by tankers) comes from a bilateral monopoly transaction. Indeed, the pipeline structure imposes a long term constraint, while trade by tankers is a lot more competitive (yet often not economically viable, as discussed above).

\clearpage

\section{Descriptive summary statistics}\label{App_desc_stat}

In the current Appendix we display below the descriptive summary statistics for pipeline connections in Table \ref{tab:pipeline_distribution}, followed further below in Table \ref{desc_stat_gen} by general summary statistics of all main variables used in the statistical analysis.

\begin{table}[htbp]\centering
\caption{Distribution of international natural gas pipeline connections}\label{tab:pipeline_distribution}
\begin{threeparttable}
\begin{tabular}{l | c c c c|}
\hline\hline
    & 1993 & 2000 &  2010 & 2018\\
\hline
Within Africa &1&2 & 6 & 7\\
Within Asia&12 & 14 & 27 & 29\\
Within Europe	&36 &48	&53 & 64\\
\hline
Africa-Asia& 0 & 0 & 2 & 3\\
Africa-Europe&2&3 & 5 & 5\\
Asia-Europe&9&9 & 13 & 14\\
\hline
\hline
\end{tabular}
\begin{tablenotes}
\item \footnotesize \emph{Note:} This table displays the evolution of the interconnection of natural gas pipelines between and within continents using our novel dataset. The numbers correspond to the number of countries connected, not the number of pipelines. Russia is counted as part of Asia.
\end{tablenotes}
\end{threeparttable}
\end{table}

\begin{table}[htbp]\centering
\footnotesize
\caption{Summary statistics}\label{desc_stat_gen}
\begin{threeparttable}
\begin{tabular}{l*{1}{cccccc}}

\hline\hline

                    &        mean&          sd&         p50&         min&         max&       count\\
\hline
\textbf{Outcomes: Leader turnover}     &           &            &            &            &            &            \\
Any change in executive head of gov&       0.205&       0.404&       0.000&       0.000&       1.000&        1862\\
Leader changed (Archigos)&       0.166&       0.372&       0.000&       0.000&       1.000&        1260\\
\textbf{Outcomes: Democracy indices}     &           &            &            &            &            &            \\
Polity IV index            &       3.129&       6.547&       6.000&     -10.000&      10.000&        2900\\
Deliberative democracy index (Vdem)&       0.411&       0.260&       0.359&       0.010&       0.883&        3098\\
Level of Democracy (Freedom House)&       6.008&       3.174&       6.500&       0.000&      10.000&        2941\\
\textbf{Outcomes: Sanctions}     &           &            &            &            &            &            \\
Any sanct. imp. by cou. within netw.        &       0.088&       0.283&       0.000&       0.000&       1.000 &        3173\\
Any sanct. imp. by cou. outside netw.        &       0.411&       0.492&       0.000&       0.000&       1.000&        3173\\
Arms sanct. imp. by cou. within netw.          &       0.054&       0.226&       0.000&       0.000&       1.000&        3173\\
Arms sanct. imp. by cou. outs. netw.     &       0.138&       0.345&       0.000&       0.000&       1.000&        3173\\
Sanct. prev. war by cou. with. netw.    &       0.028&       0.164&       0.000&       0.000&       1.000&        3173\\
Sanct. prev. war by cou. outs. netw.  &       0.112&       0.315&       0.000&       0.000&       1.000&        3173\\
\textbf{Other outcomes}     &           &            &            &            &            &            \\
ln(GDP  pc ppp (in constant USD))             &       9.106&       1.289&       9.213&       6.151&      11.701&        3173\\
ICRG Indicator of Quality of Gov.&       0.572&       0.209&       0.556&       0.111&       1.000&        2415\\
Hanson \& Sigman State Capac. Index&       0.520&       0.966&       0.379&      -1.744&       2.964&        2588\\
\textbf{Explanatory var.: Centrality meas.}     &           &            &            &            &            &            \\
DC                  &       1.280&       1.919&       0.000&       0.000&      12.000&        3173\\
Betweenness centrality&       0.013&       0.043&       0.000&       0.000&       0.444&        3173\\
Ln(BC)              &      -7.445&       2.162&      -8.567&      -8.567&      -0.812&        3173\\
\textbf{Control variables}     &           &            &            &            &            &            \\
Population          &43186024.581&   1.560e+08& 9148092.000&   74205.000&   1.403e+09&        3173\\
lnpop               &      16.014&       1.712&      16.029&      11.215&      21.062&        3173\\
GDP pc ppp (in constant USD)&   18356.598&   20956.731&   10026.582&     469.190&  120647.820&        3173\\
Oil rents (\% of GDP)&       4.084&      10.223&       0.010&       0.000&      66.713&        3173\\
Log(Oil rents (\% of GDP))&       4.084&      10.223&       0.010&       0.000&      66.713&        3173\\
Natural gas rents (\% of GDP)&       0.723&       3.347&       0.001&       0.000&      68.564&        3173\\
Log(Natural gas rents (\% of GDP))&       0.723&       3.347&       0.001&       0.000&      68.564&        3173\\
\hline\hline
\end{tabular}
\begin{tablenotes}
\item \footnotesize \emph{Note:} Data sources listed and discussed in Section \ref{Sec_data} of the main text.
\end{tablenotes}
\end{threeparttable}
\end{table}

%
%

\clearpage

\section{Regression tables}\label{app:tables}

In what follows, we shall display the full regression tables corresponding to the key results of the paper in the specifications with the full battery of controls, represented graphically in the main text.

\begin{table}[htbp]\centering
\def\sym#1{\ifmmode^{#1}\else\(^{#1}\)\fi}
\caption{Leader durability}\label{table_leaderchange}
\begin{threeparttable}
\begin{tabular}{l*{2}{c}}
\hline\hline
                    &(1)& (2)\\
\textbf{Dataset: }         &Democracy and Dictatorship & Archigos\\
\textbf{Dep. Var.: Leader turnover }  & & \\
\hline
$ln(BC_{it-1})$            &    -0.05\sym{***}&    -0.06\sym{**} \\
                    &   [0.02]         &   [0.02]         \\
[1em]
$DC_{it-1}$                   &     0.04         &     0.02         \\
                    &   [0.04]         &   [0.06]         \\
\hline
\(R^{2}\)           &    0.210         &    0.224         \\
Mean dep var        &     0.20         &     0.17         \\
FE year             &      Yes         &      Yes         \\
FE ctry             &      Yes         &      Yes         \\
Controls             &      Yes         &      Yes   \\
Observations        &     1862         &     1260         \\
\hline\hline
\end{tabular}
\begin{tablenotes}
\item \footnotesize  \emph{Note: }\sym{*} \(p<0.10\), \sym{**} \(p<0.05\), \sym{***} \(p<0.01\). Linear Probability Models estimated with OLS. Standard errors clustered at the country level. The unit of observation is country-year and the dataset spans from 1993 to 2018 covering Africa, Asia and Europe. The outcome is a dummy taking the value one if the executive leader changed during the year. $ln(BC_{it-1})$ represents the lagged natural logarithm of the natural gas pipeline betweenness centrality for country $i$.  $DC_{it-1}$ represents the lagged natural gas pipeline degree centrality for country $i$. The vector of controls includes: the log of the population, the log of the GDP per capita in ppp, the log of natural gas rents as well as the log of oil rents both in percent of the GDP.
\end{tablenotes}
\end{threeparttable}
\end{table}

\clearpage

\begin{table}[htbp]\centering
\def\sym#1{\ifmmode^{#1}\else\(^{#1}\)\fi}
\caption{Sanctions}\label{table_sanctions}
\begin{threeparttable}
\begin{tabular}{l c c | c c | c c}
\hline\hline

                    &\multicolumn{1}{c}{(1)}&\multicolumn{1}{c}{(2)}&\multicolumn{1}{c}{(3)}&\multicolumn{1}{c}{(4)}&\multicolumn{1}{c}{(5)}&\multicolumn{1}{c}{(6)}\\
                    &\multicolumn{2}{c}{Any Sanction         }                 & \multicolumn{2}{c}{Arms Sanctions        }                &\multicolumn{2}{c}{Sanctions to prevent war      }                  \\
                    &\multicolumn{1}{c}{Network}&\multicolumn{1}{c}{Others}&\multicolumn{1}{c}{Network}&\multicolumn{1}{c}{Others}&\multicolumn{1}{c}{Network}&\multicolumn{1}{c}{Others}\\
\hline
$ln(BC_{it-1})$          &    -0.03         &     0.02         &    -0.04\sym{**} &     0.01         &    -0.03\sym{**} &     0.01         \\
                    &   [0.02]         &   [0.02]         &   [0.02]         &   [0.02]         &   [0.01]         &   [0.01]         \\
[1em]
$DC_{it-1}$                 &     0.09\sym{**} &    -0.10\sym{***}&     0.10\sym{***}&    -0.05         &     0.07\sym{**} &    -0.06\sym{**} \\
                    &   [0.03]         &   [0.04]         &   [0.03]         &   [0.04]         &   [0.03]         &   [0.02]         \\
[1em]
\hline
\(R^{2}\)           &    0.543         &    0.580         &    0.582         &    0.599         &    0.526         &    0.623         \\
Mean dep var        &    0.088         &    0.411         &    0.054         &    0.138         &    0.028         &    0.112         \\
FE year             &      Yes         &      Yes         &      Yes         &      Yes         &      Yes         &      Yes         \\
FE ctry             &      Yes         &      Yes         &      Yes         &      Yes         &      Yes         &      Yes         \\
Controls    &      Yes         &      Yes         &      Yes         &      Yes         &      Yes         &      Yes         \\
Observations        &     3173         &     3173         &     3173         &     3173         &     3173         &     3173         \\
\hline\hline
\end{tabular}
\begin{tablenotes}
\item \footnotesize \emph{Note:}  \sym{*} \(p<0.10\), \sym{**} \(p<0.05\), \sym{***} \(p<0.01\). Standard errors clustered at the country level. Linear Probability Models estimated with OLS. The unit of observation is country-year and the dataset spans from 1993 to 2018 covering Africa, Asia and Europe. The outcome is a dummy taking the value one if country $i$ was subject to sanctions in year $t$. There are two dimensions with respect to the outcome. First, we use three different types of sanctions: any type of sanctions (models (1)-(2)); then arms sanctions (models (3)-(4)); and finally sanctions to prevent war (models (5)-(6)). Second, we restrict attention to different sets of countries imposing the sanctions: sanctions imposed by countries connected to country $i$ in the natural gas pipeline network (models (1), (3), and (5));
     then in (2), (4) and (6) we restrict the analysis to sanctions imposed by countries not connected to country $i$ within the pipeline network. $ln(BC_{it-1})$ represents the lagged natural logarithm of the natural gas pipeline betweenness centrality for country $i$.  $DC_{it-1}$ represents the lagged natural gas pipeline degree centrality for country $i$. The vector of controls includes: the log of the population, the log of the GDP per capita in ppp, the log of natural gas rents as well as the log of oil rents both in percent of the GDP.
\end{tablenotes}
\end{threeparttable}
\end{table}

\clearpage

\begin{table}
\footnotesize
\def\sym#1{\ifmmode^{#1}\else\(^{#1}\)\fi}
\caption{Democracy}\label{table_democracy}
\begin{threeparttable}
\begin{tabular}{l*{6}{c}}
\hline\hline
        &\multicolumn{1}{c}{(1)}&\multicolumn{1}{c}{(2)}&\multicolumn{1}{c}{(3)}\\
\textbf{Dataset: }                               &\multicolumn{1}{c}{PolityIV}&\multicolumn{1}{c}{V-dem}&\multicolumn{1}{c}{Freedom House}\\
\textbf{Dep. Var.:   }                    &Democracy & Democracy & Democracy \\
\hline
$ln(BC_{it-1})$             &    -0.12         &    -0.01         &    -0.06         \\
                    &   [0.63]         &   [0.45]         &   [0.55]         \\
[1em]
$DC_{it-1}$          &    -0.27         &    -0.01         &    -0.02         \\
                    &   [0.48]         &   [0.48]         &   [0.88]         \\
\hline
\(R^{2}\)           &    0.888         &    0.936         &    0.922         \\
Mean dep var        &     3.13         &     0.41         &     6.01         \\
FE year             &      Yes         &      Yes         &      Yes         \\
FE ctry             &      Yes         &      Yes         &      Yes         \\
Observations        &     2899         &     3098         &     2940         \\
\hline\hline
\hline\hline
\end{tabular}
\begin{tablenotes}
\item \footnotesize  \sym{*} \(p<0.10\), \sym{**} \(p<0.05\), \sym{***} \(p<0.01\). Standard errors clustered at the country level. The unit of observation is country-year and the dataset spans from 1993 to 2018 covering Africa, Asia and Europe. The outcomes are different democracy scales: model (1) uses PolityIV a scale from -10 (total autocracy) to 10 (perfect democracy) \citep{polity5}, model (2) uses V-dem an index from 0 to 1 quantifying the extent to which the ideal of deliberative democracy is achieved \citep{pemstein2018v}, and model (3) uses a scale from 0 to 10 (perfect democracy) from \cite{house2022freedom}. $ln(BC_{it-1})$ represents the lagged natural logarithm of the natural gas pipeline betweenness centrality for country $i$.  $DC_{it-1}$ represents the lagged natural gas pipeline degree centrality for country $i$. The vector of controls includes: the log of the GDP (except for the regression with GDP as the outcome), the log of the population, the log of natural gas rents as well as the log of oil rents both in percent of the GDP.
\end{tablenotes}
\end{threeparttable}
\end{table}

\clearpage

\begin{table}
\footnotesize
\def\sym#1{\ifmmode^{#1}\else\(^{#1}\)\fi}
\caption{Growth, state capacity, and quality of the government}\label{table_goodgas}
\begin{threeparttable}
\begin{tabular}{l*{6}{c}}
\hline\hline
       &\multicolumn{1}{c}{(1)}&\multicolumn{1}{c}{(2)}&\multicolumn{1}{c}{(3)}\\
\textbf{Dataset: }                             &\multicolumn{1}{c}{World Bank}&\multicolumn{1}{c}{ICRG}&\multicolumn{1}{c}{\citep{hanson2021leviathan}}\\
\textbf{Dep. Var.:   }                & ln(GDP)  & Quality of the Gov. & State Capacity   \\
\hline
$ln(BC_{it-1})$         &     0.01         &     0.00         &    -0.00         \\
                    &   [0.37]         &   [0.25]         &   [0.79]         \\
[1em]
$DC_{it-1}$          &    -0.01         &    -0.01\sym{**} &     0.01         \\
                    &   [0.58]         &   [0.02]         &   [0.59]         \\
\hline
\(R^{2}\)           &    0.983         &    0.955         &    0.967         \\
Mean dep var        &     0.04         &     0.57         &     0.52         \\
FE year             &      Yes         &      Yes         &      Yes         \\
FE ctry             &      Yes         &      Yes         &      Yes         \\
Observations        &     3173         &     2415         &     2588         \\
\hline\hline
\hline\hline
\end{tabular}
\begin{tablenotes}
\item \footnotesize  \sym{*} \(p<0.10\), \sym{**} \(p<0.05\), \sym{***} \(p<0.01\). Standard errors clustered at the country level. The unit of observation is country-year and the dataset spans from 1993 to 2018 covering Africa, Asia and Europe. The outcome for model (1) is the log of the GDP per capita in ppp. The outcome for model (2) is a measure of the quality of the government from the International Country Risk Guide based on “Corruption”, “Law and Order” and “Bureaucracy Quality” going from 0 to 1 (with higher value indicating better quality). The outcome for model (3) is a State Capacity Index going from -3 (worst) to 3 (best) \citep{hanson2021leviathan}. $ln(BC_{it-1})$ represents the lagged natural logarithm of the natural gas pipeline betweenness centrality for country $i$.  $DC_{it-1}$ represents the lagged natural gas pipeline degree centrality for country $i$. The vector of controls includes: the log of the GDP (except for the regression with GDP as the outcome), the log of the population, the log of natural gas rents as well as the log of oil rents both in \% of the GDP.
\end{tablenotes}
\end{threeparttable}
\end{table}

\clearpage

\section{Robustness}\label{app:robustness}

This Appendix is devoted to several sensitivity tests. In particular, we investigate sensitivity to time lags, internal conflict and outlier removal.

\subsection{Lags}

Below we shall display estimates for a different time lag structure. In the main baseline regressions we focus on the first lag, which reflects a reasonable time assumption for deploying effects and attenuates concerns about reversed causality. Here we also show estimates for t-2 and contemporary (t) effects. We find that indeed the effect on regime survival arises in t-1, while for sanctions the effects span over different years and for democracy and other outcome variables no effect is found for any lag structure.

\begin{figure}[ht!]
    \centering
    \begin{subfigure}[t]{0.49\textwidth}
        \centering
    	\includegraphics[width=\textwidth]{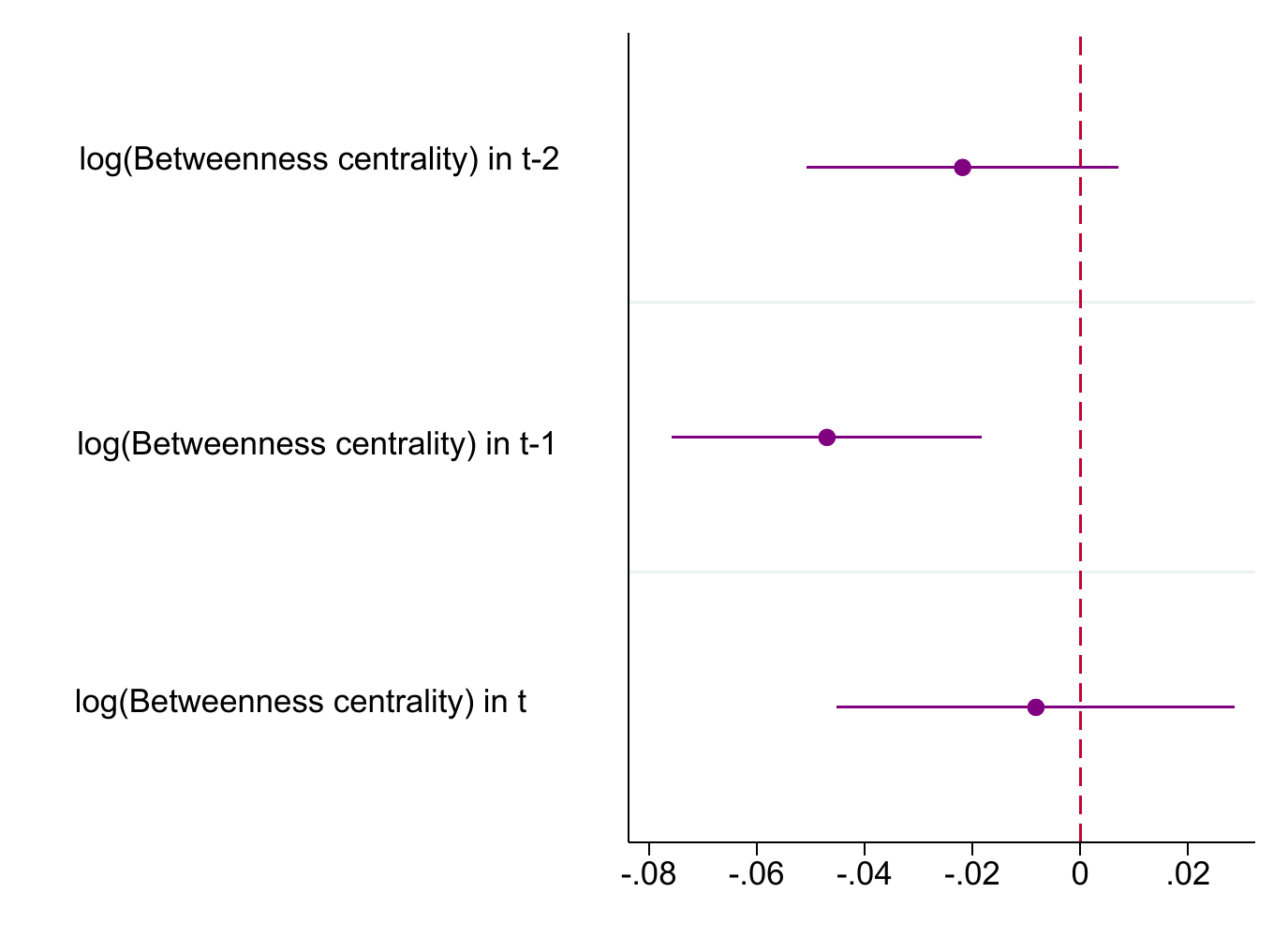}
        \caption{Democracy and dictatorship dataset}
    \end{subfigure}
    \begin{subfigure}[t]{0.49\textwidth}
        \centering
         \includegraphics[width=\textwidth]{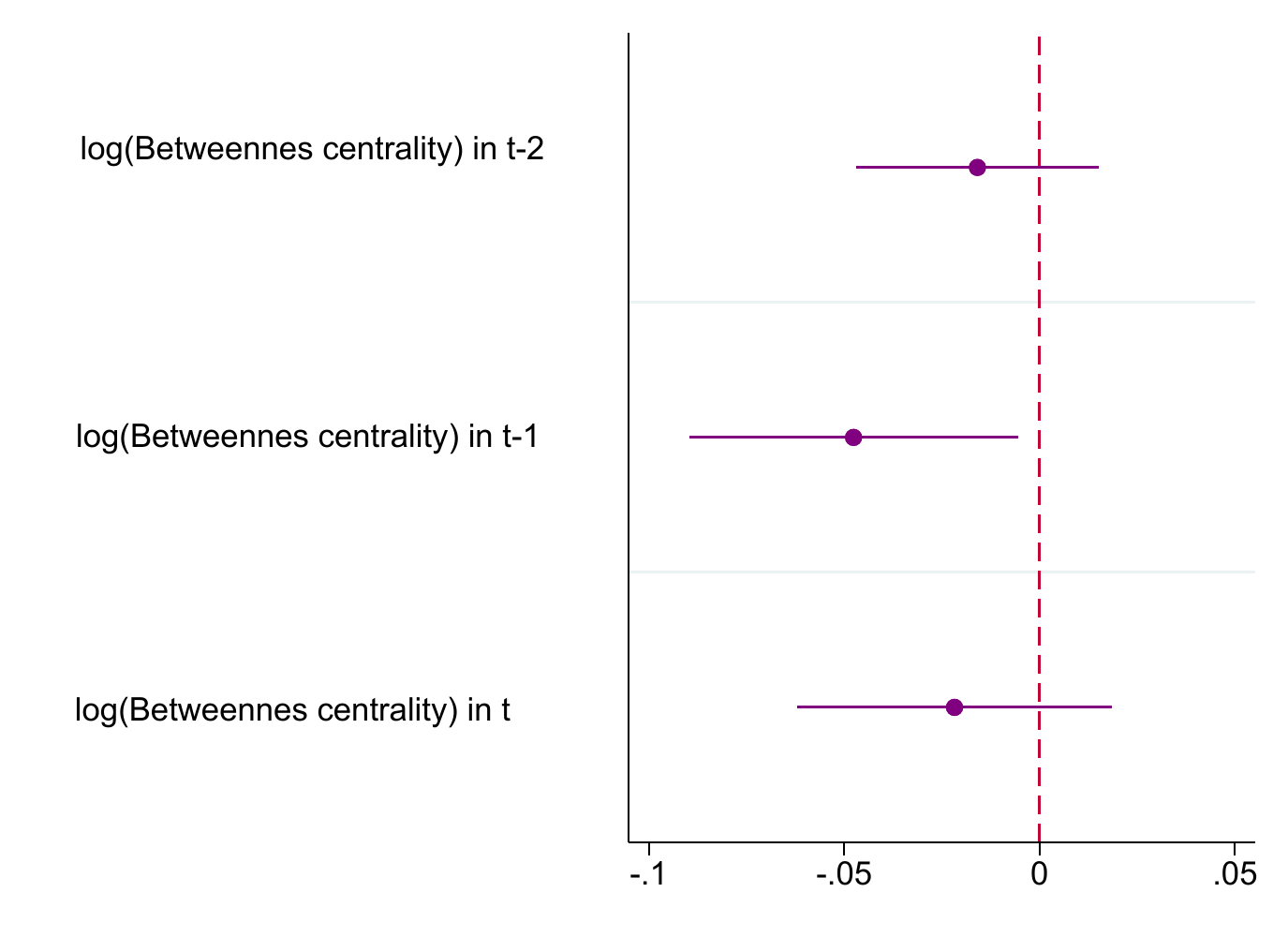}
        \caption{Archigos dataset}
    \end{subfigure}
	\caption{Leaders turnover: Different time lags}
 \footnotesize  \emph{Note: }Value and 95\% confidence intervals for the betweenness centrality coefficient computed for models using different lags. Linear Probability Models estimated with OLS. Standard error clustered at the country level. The unit of observation is country-year and the dataset spans from 1993 to 2018 covering Africa, Asia and Europe. The outcome is a dummy taking the value one if the executive leader changed during the year. Betweenness centrality corresponds to the natural logarithm of the natural gas pipeline betweenness centrality for country $i$. The vector of controls includes: degree centrality for country $i$, the log of the population, the log of the GDP per capita in ppp, the log of natural gas rents as well as the log of oil rents both in percent of the GDP.
\end{figure}

\begin{figure}[ht!]
    \centering
    \begin{subfigure}[t]{0.49\textwidth}
        \centering
    	\includegraphics[width=\textwidth]{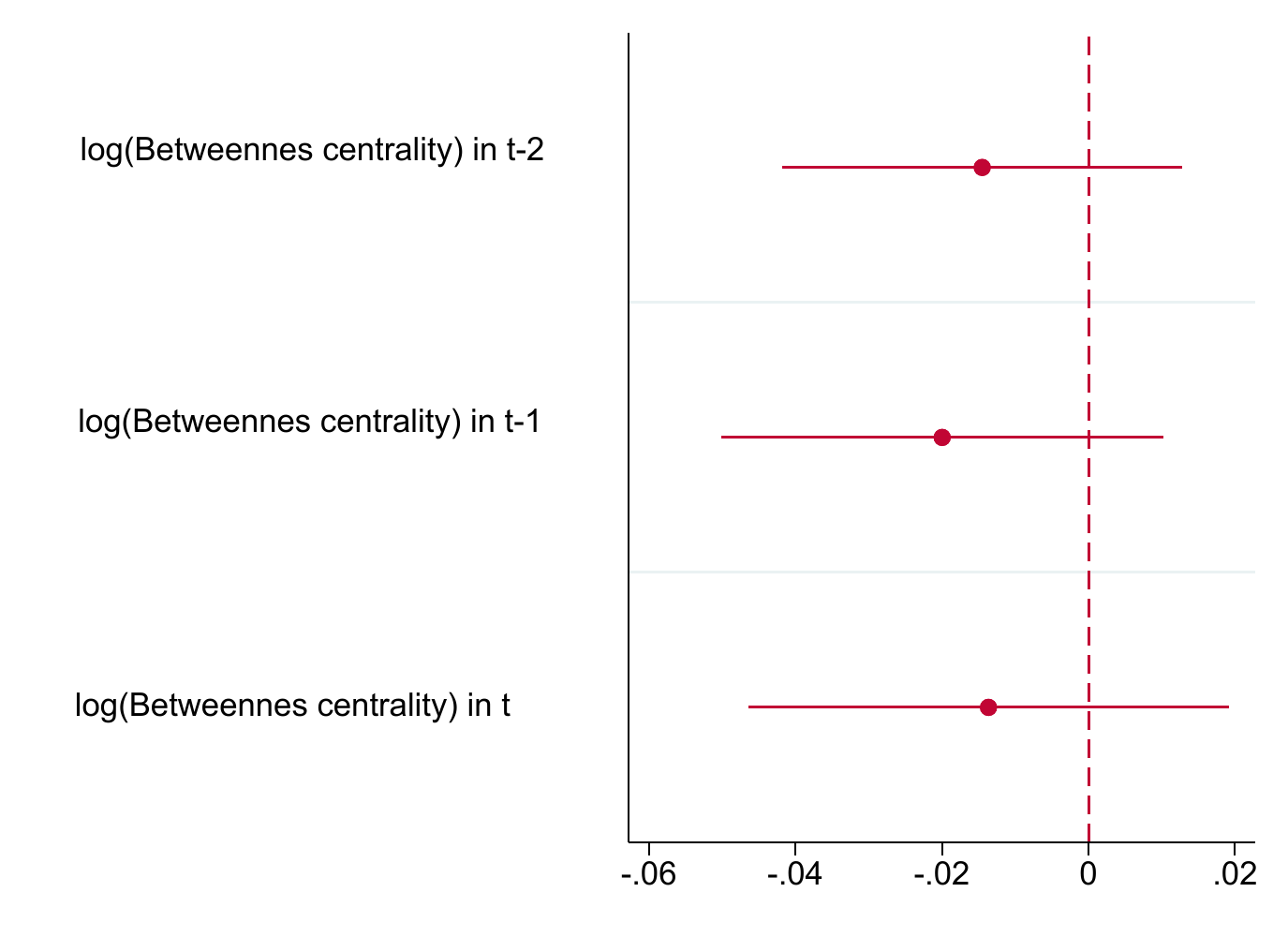}
        \caption{Any sanctions}
    \end{subfigure}
    \begin{subfigure}[t]{0.49\textwidth}
        \centering
         \includegraphics[width=\textwidth]{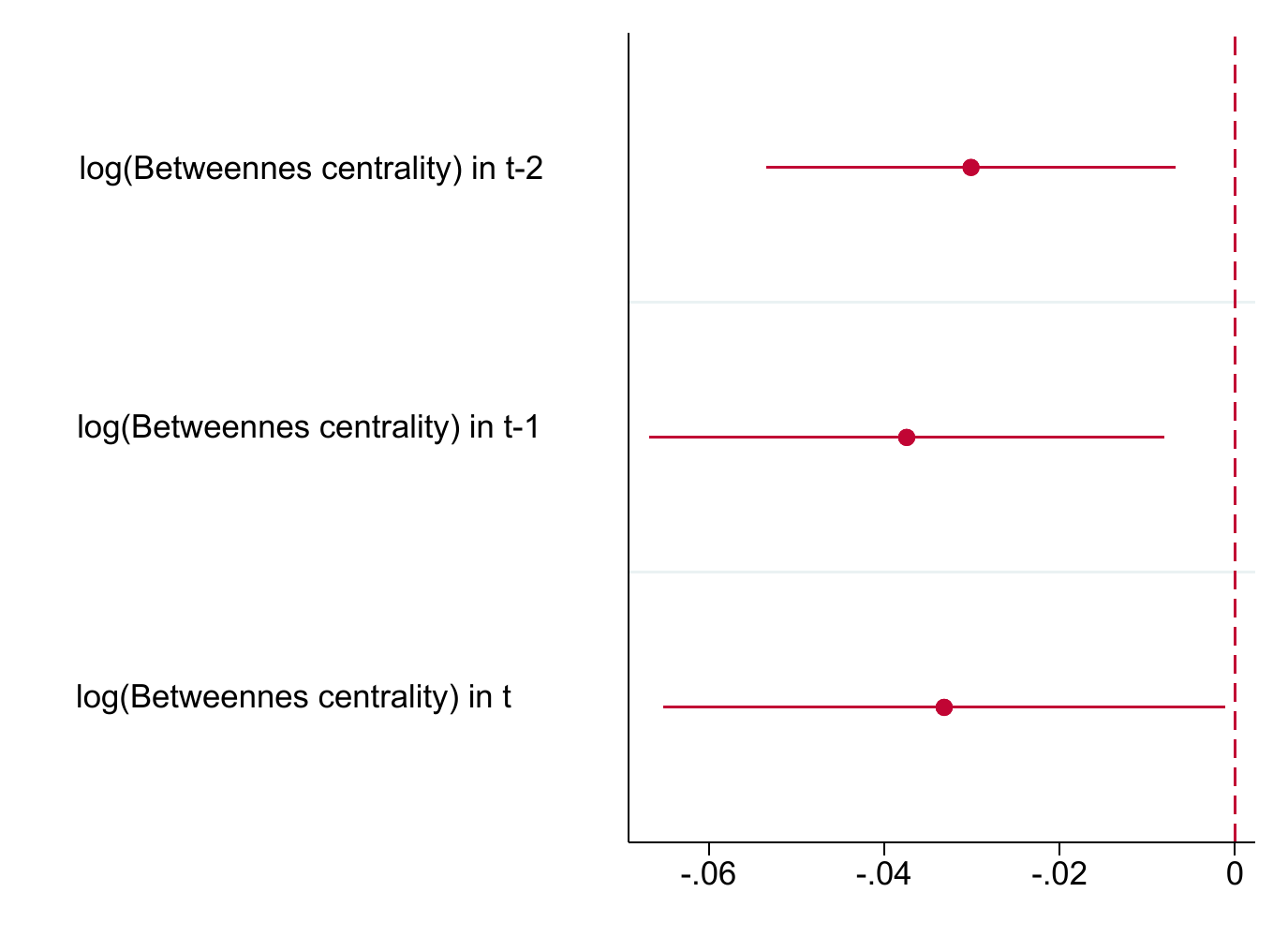}
        \caption{Arms sanctions}
    \end{subfigure}\\
    \begin{subfigure}[t]{0.49\textwidth}
        \centering
         \includegraphics[width=\textwidth]{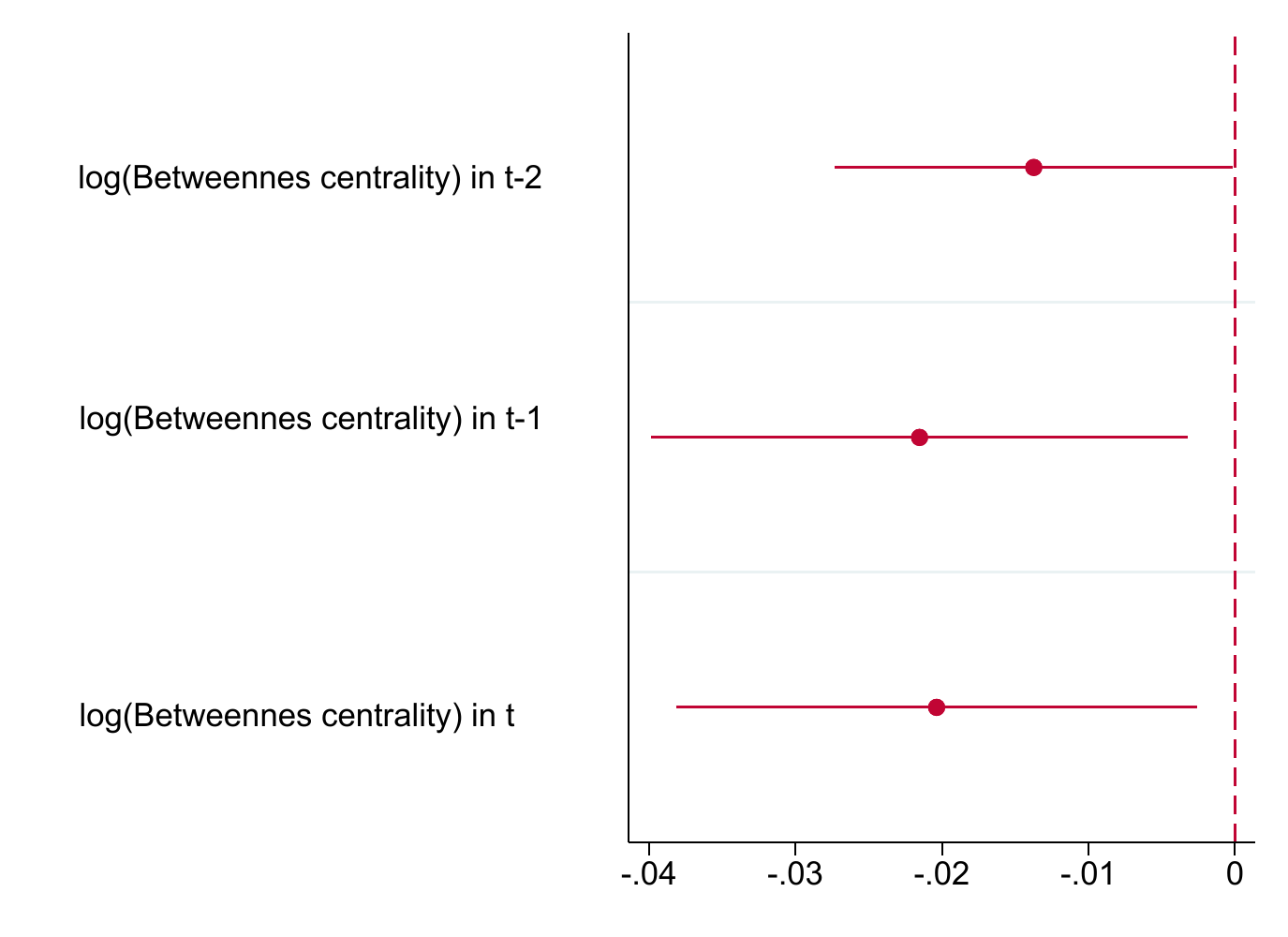}
        \caption{Sanctions to prevent war}
    \end{subfigure}
	\caption{Sanctions: Different time lags.}
 \footnotesize  \emph{Note: }Value and 95\% confidence intervals for the betweenness centrality coefficient computed for models using different lags. Linear Probability Models estimated with OLS. Standard error clustered at the country level. The unit of observation is country-year and the dataset spans from 1993 to 2018 covering Africa, Asia and Europe. The outcome is a dummy taking the value one if country $i$ was subject to sanctions in year $t$. We use three different types of sanctions: any type of sanctions; then arms sanctions; and finally sanctions to prevent war. Betweenness centrality corresponds to the natural logarithm of the natural gas pipeline betweenness centrality for country $i$. The vector of controls includes: degree centrality for country $i$, the log of the population, the log of the GDP per capita in ppp, the log of natural gas rents as well as the log of oil rents both in percent of the GDP.
\end{figure}

\begin{figure}[ht!]
    \centering
    \begin{subfigure}[t]{0.49\textwidth}
        \centering
    	\includegraphics[width=\textwidth]{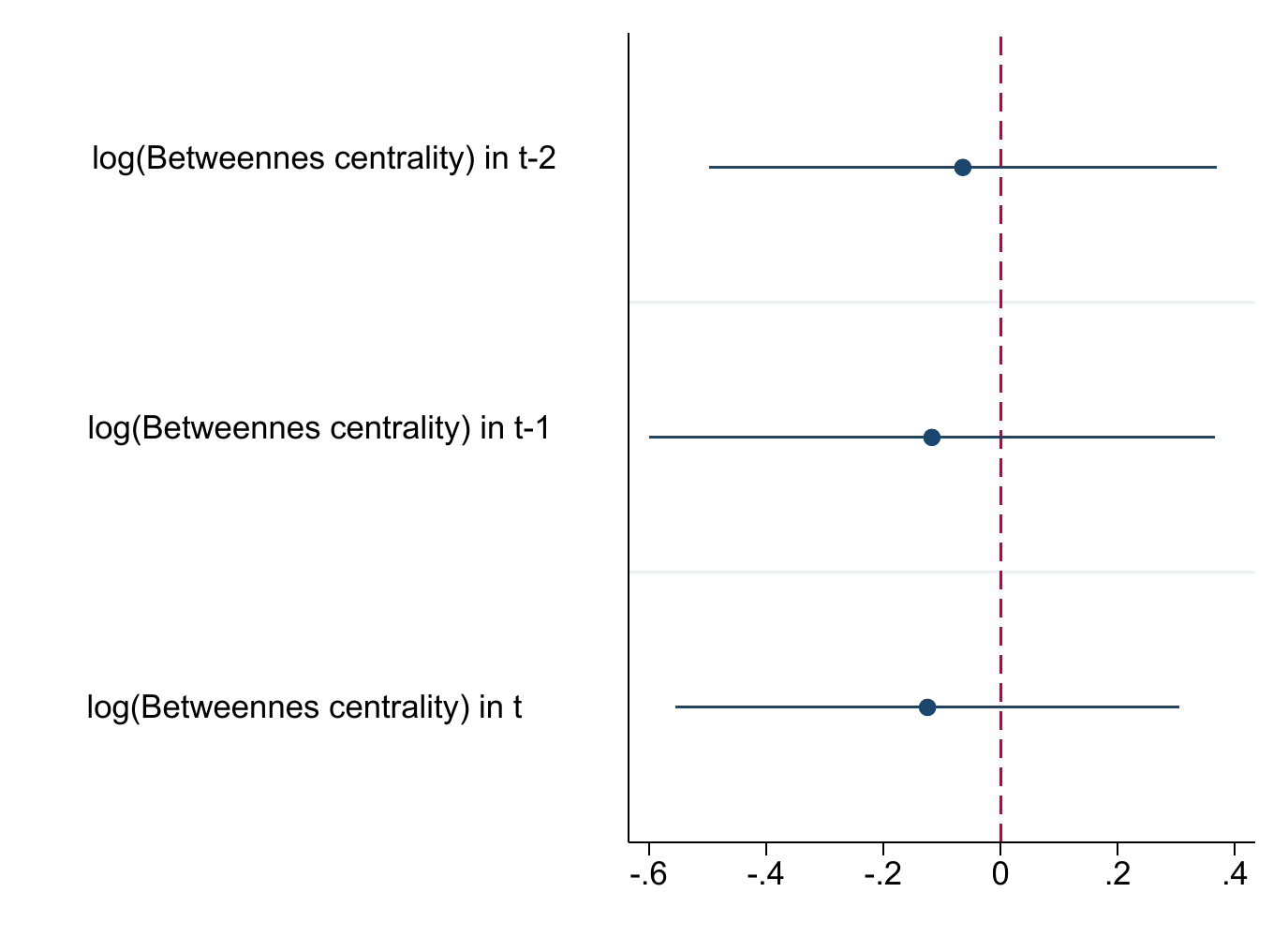}
        \caption{Polity IV dataset}
    \end{subfigure}
    \begin{subfigure}[t]{0.49\textwidth}
        \centering
         \includegraphics[width=\textwidth]{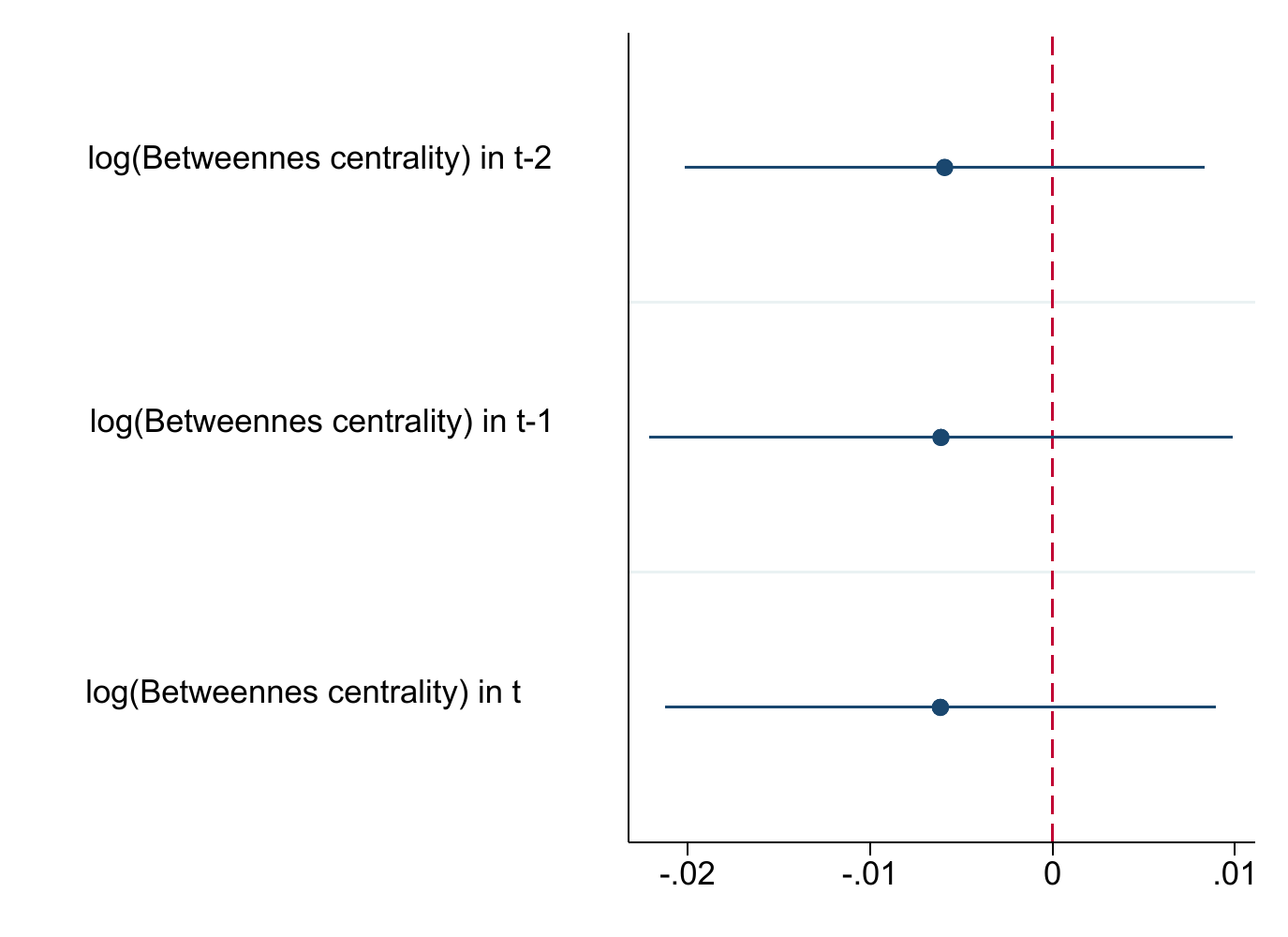}
        \caption{V-Dem dataset}
    \end{subfigure}\\
    \begin{subfigure}[t]{0.49\textwidth}
        \centering
         \includegraphics[width=\textwidth]{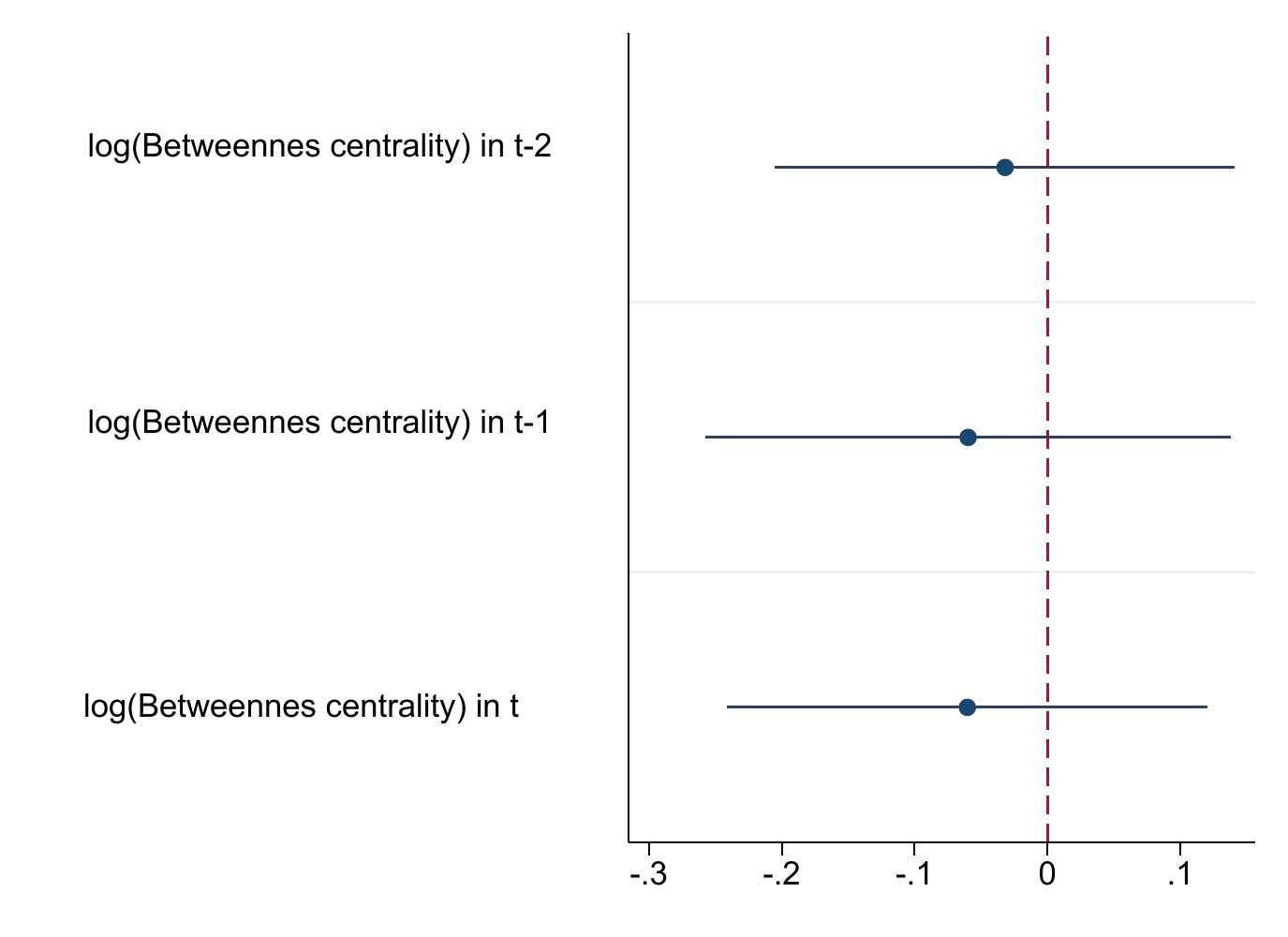}
        \caption{Freedom House dataset}
    \end{subfigure}
	\caption{Democracy: Different time lags}
\footnotesize   Value and 95\% confidence intervals for the betweenness centrality coefficient computed for models using different lags. Linear Probability Models estimated with OLS. Standard error clustered at the country level. The unit of observation is country-year and the dataset spans from 1993 to 2018 covering Africa, Asia and Europe. The outcomes are different democracy scales: model (1) uses Polity IV a scale from -10 (total autocracy) to 10 (perfect democracy) \citep{polity5}, model (2) uses V-dem an index from 0 to 1 quantifying the extent to which the ideal of deliberative democracy is achieved \citep{pemstein2018v}, and model (3) uses a scale from 0 to 10 (perfect democracy) from \cite{house2022freedom}. $ln(BC_{it-1})$ represents the lagged natural logarithm of the natural gas pipeline betweenness centrality for country $i$. The vector of controls includes: $DC_{it-1}$ which represents the lagged natural gas pipeline degree centrality for country $i$, the log of the GDP (except for the regression with GDP as the outcome), the log of the population, the log of natural gas rents as well as the log of oil rents both in percent of the GDP.
\end{figure}

\begin{figure}[ht!]
    \centering
    \begin{subfigure}[t]{0.49\textwidth}
        \centering
    	\includegraphics[width=\textwidth]{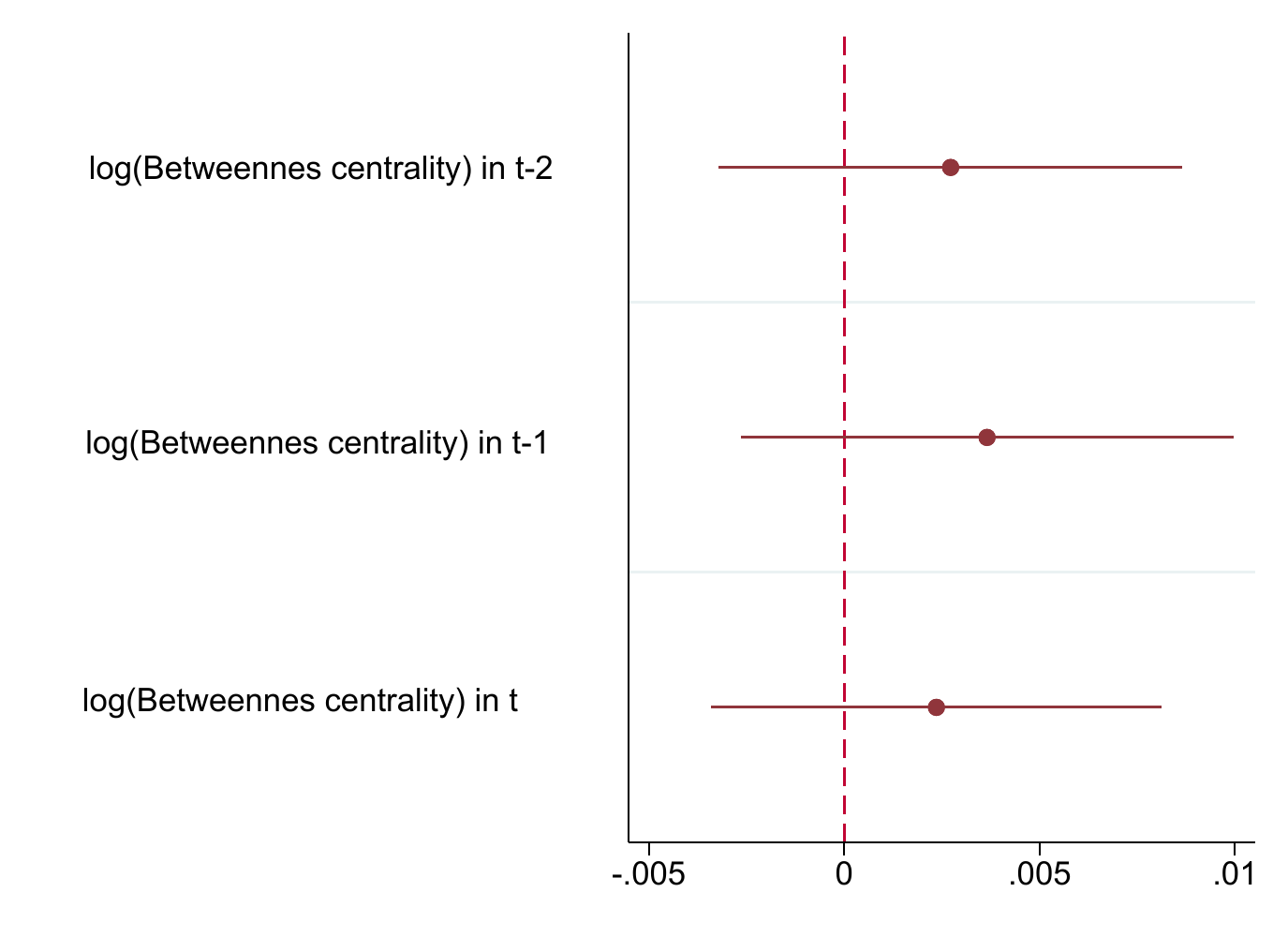}
        \caption{log(GDP per capita in p.p.p)}
    \end{subfigure}
    \begin{subfigure}[t]{0.49\textwidth}
        \centering
         \includegraphics[width=\textwidth]{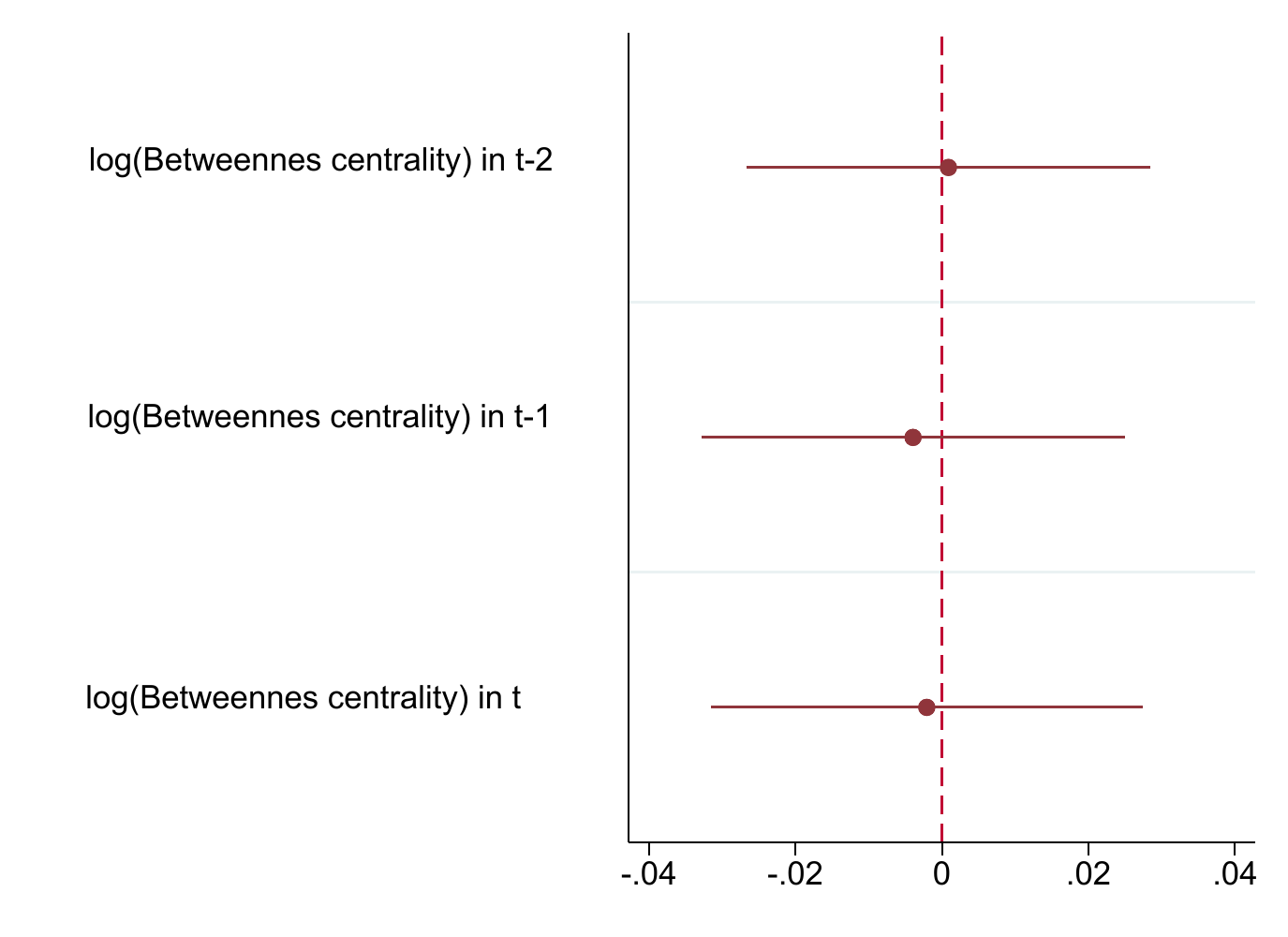}
        \caption{State Capacity (ICRG dataset)}
    \end{subfigure}\\
    \begin{subfigure}[t]{0.49\textwidth}
        \centering
         \includegraphics[width=\textwidth]{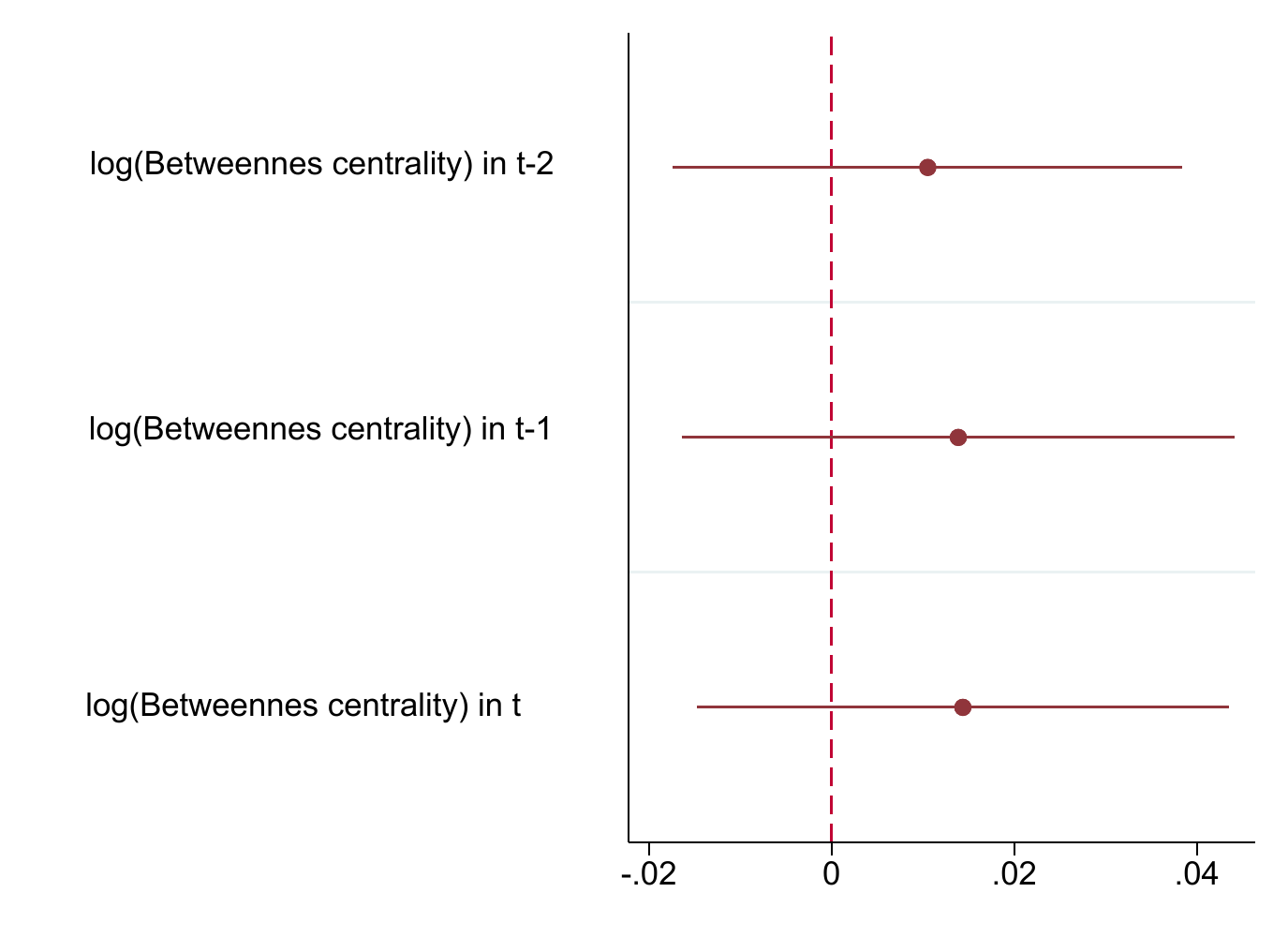}
        \caption{Quality of the government (Hanson and Sigman (2021) dataset)}
    \end{subfigure}
	\caption{Other outcomes: Different time lags}
\footnotesize  Value and 95\% confidence intervals for the betweenness centrality coefficient computed for models using different lags. Linear Probability Models estimated with OLS. Standard errors clustered at the country level. The unit of observation is country-year and the dataset spans from 1993 to 2018 covering Africa, Asia and Europe. The outcome for model (1) is the log of the GDP per capita in ppp. The outcome for model (2) is a measure of the quality of the government from the International Country Risk Guide based on “Corruption”, “Law and Order” and “Bureaucracy Quality” going from 0 to 1 (with higher value indicating better quality). The outcome for model (3) is a State Capacity Index going from -3 (worst) to 3 (best) \citep{hanson2021leviathan}. $ln(BC_{it-1})$ represents the lagged natural logarithm of the natural gas pipeline betweenness centrality for country $i$.  The vector of controls includes: $DC_{it-1}$ which represents the lagged natural gas pipeline degree centrality for country $i$, the log of the GDP (except for the regression with GDP as the outcome), the log of the population, the log of natural gas rents as well as the log of oil rents both in \% of the GDP.
\end{figure}

\clearpage

\subsection{Controlling for internal conflicts}
\begin{figure}[ht!]
    \centering
    \begin{subfigure}[t]{0.49\textwidth}
        \centering
    	\includegraphics[width=\textwidth]{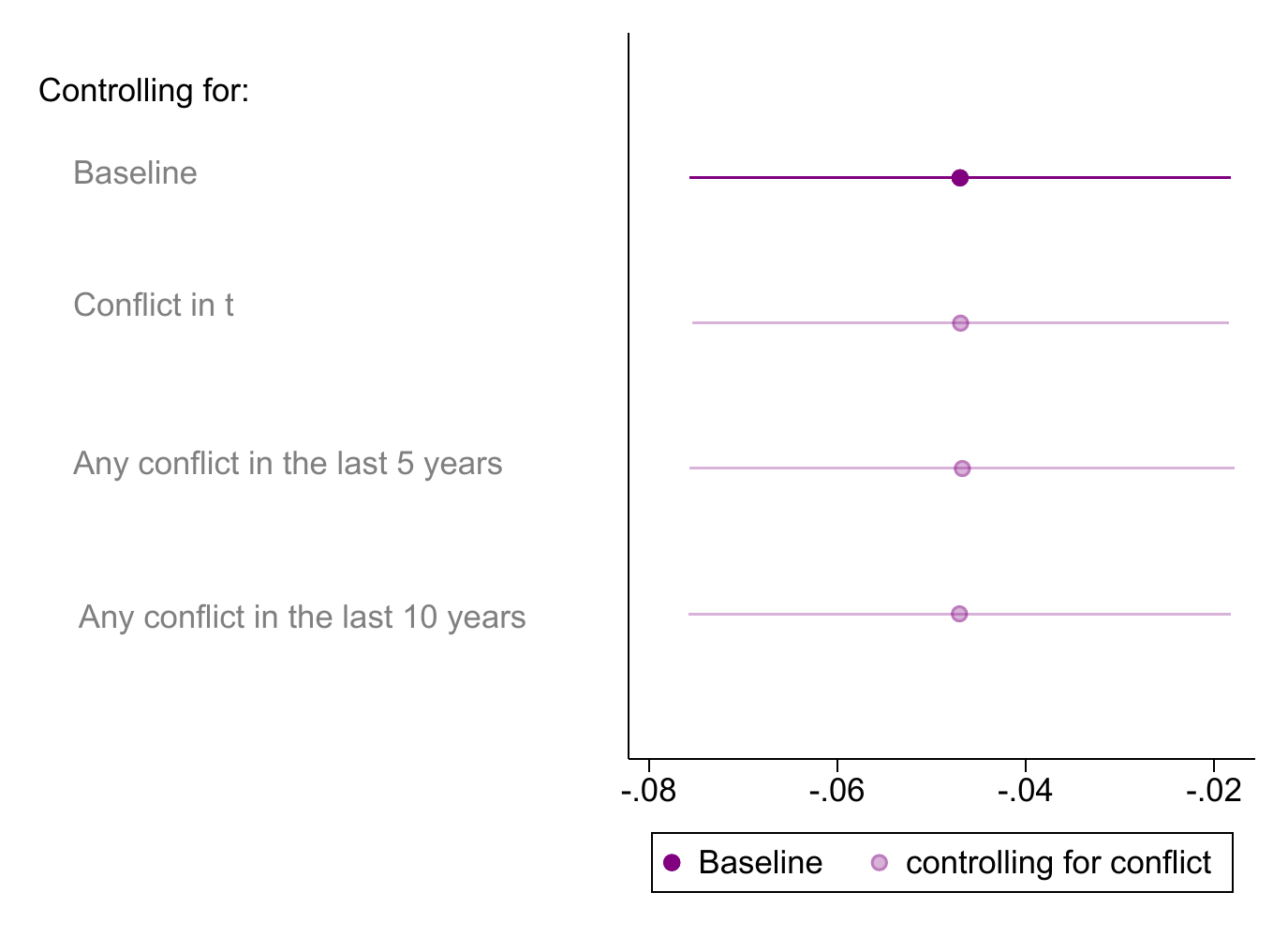}
        \caption{Democracy and dictatorship dataset}
    \end{subfigure}
    \begin{subfigure}[t]{0.49\textwidth}
        \centering
         \includegraphics[width=\textwidth]{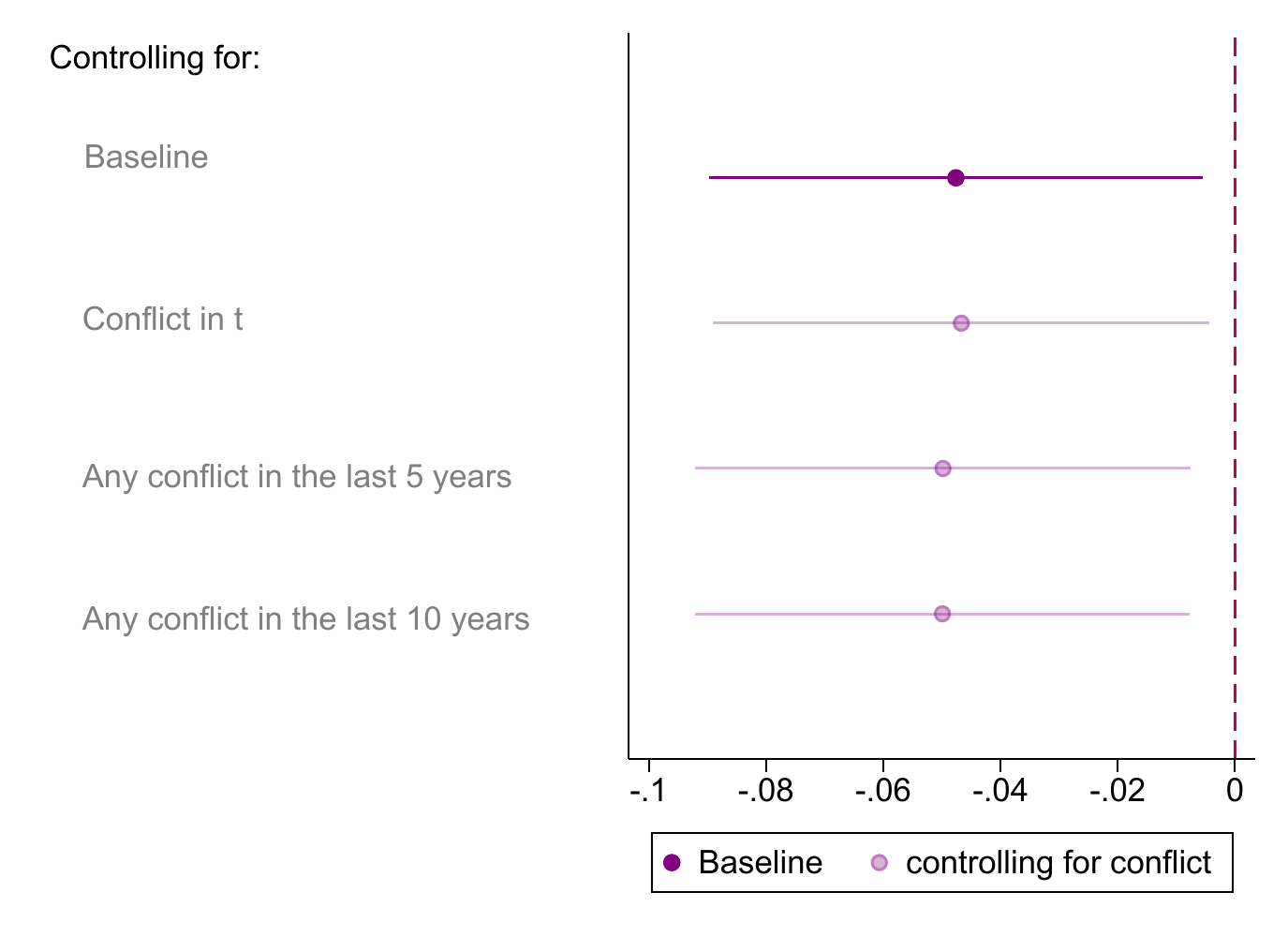}
        \caption{Archigos dataset}
    \end{subfigure}
	\caption{Leaders turnover: controlling for internal conflicts.}\label{fig:conflict_leader}
 \footnotesize  \emph{Note: }Value and 95\% confidence intervals for the betweenness centrality coefficient. Linear Probability Models estimated with OLS. Standard error clustered at the country level. The unit of observation is country-year and the dataset spans from 1993 to 2018 covering Africa, Asia and Europe. The outcome is a dummy taking the value one if the executive leader changed during the year. Betweenness centrality corresponds to the natural logarithm of the natural gas pipeline betweenness centrality for country $i$. The vector of controls includes: degree centrality for country $i$, the log of the population, the log of the GDP per capita in ppp, the log of natural gas rents as well as the log of oil rents both in percent of the GDP and internal conflict.
\end{figure}

\begin{figure}[ht!]
    \centering
    \begin{subfigure}[t]{0.49\textwidth}
        \centering
    	\includegraphics[width=\textwidth]{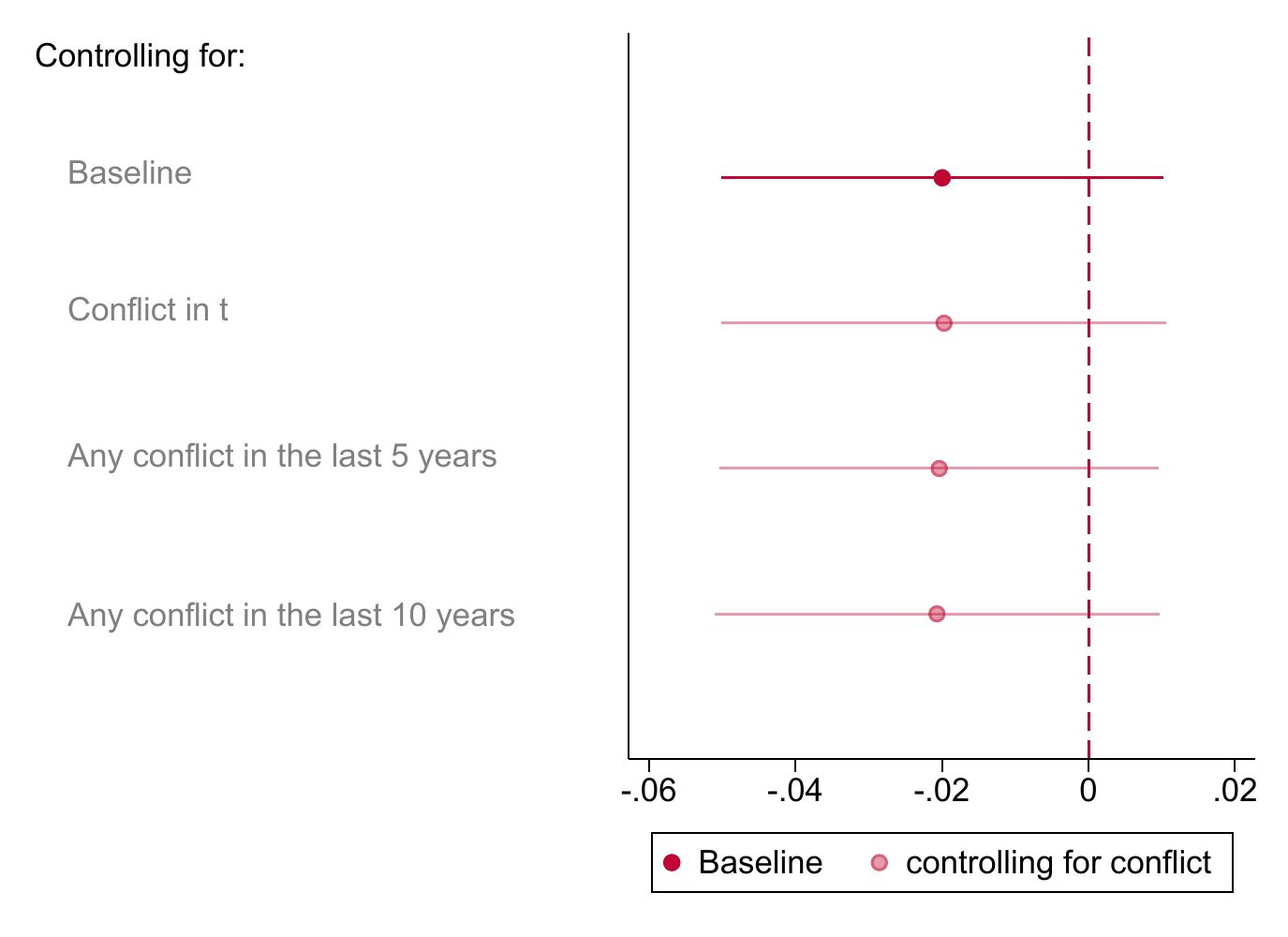}
        \caption{Any sanctions}
    \end{subfigure}
    \begin{subfigure}[t]{0.49\textwidth}
        \centering
         \includegraphics[width=\textwidth]{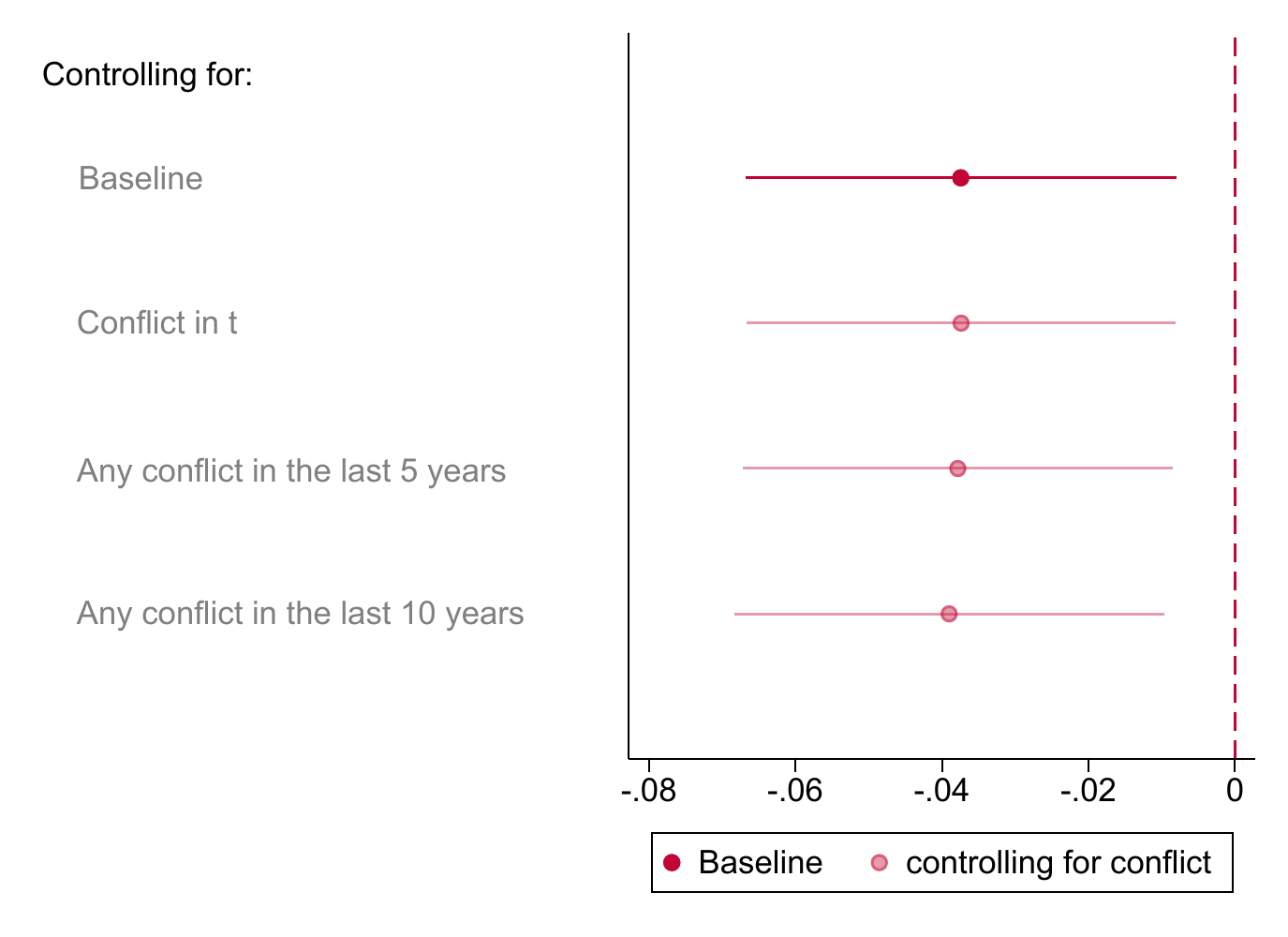}
        \caption{Arms sanctions}
    \end{subfigure}\\
    \begin{subfigure}[t]{0.49\textwidth}
        \centering
         \includegraphics[width=\textwidth]{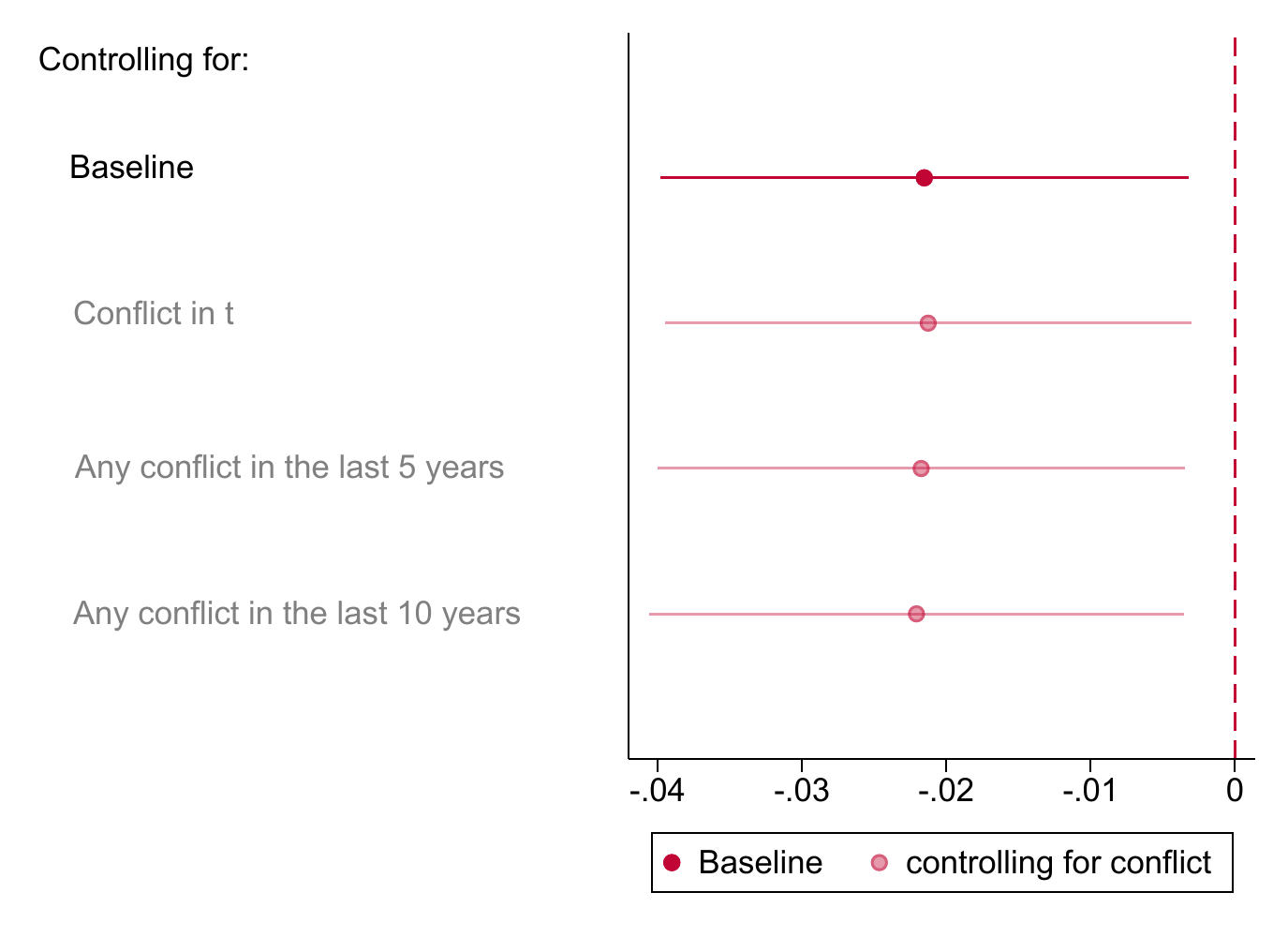}
        \caption{Sanctions to prevent war}
    \end{subfigure}
	\caption{Sanctions: controlling for internal conflicts.}\label{fig:conflict_sanction}
\footnotesize  \emph{Note: }Value and 95\% confidence intervals for the betweenness centrality coefficient. Linear Probability Models estimated with OLS. Standard error clustered at the country level. The unit of observation is country-year and the dataset spans from 1993 to 2018 covering Africa, Asia and Europe. The outcome is a dummy taking the value one if country $i$ was subject to sanctions in year $t$. We use three different types of sanctions: any type of sanctions; then arms sanctions; and finally sanctions to prevent war. Betweenness centrality corresponds to the natural logarithm of the natural gas pipeline betweenness centrality for country $i$. The vector of controls includes: degree centrality for country $i$, the log of the population, the log of the GDP per capita in ppp, the log of natural gas rents as well as the log of oil rents both in percent of the GDP and internal conflict.
\end{figure}

Below we shall display estimates for our baseline model augmented to control for contemporaneous or past internal conflicts. Internal conflict could influence the outcomes (leader turnover or sanctions) as well as the probability of building new pipelines connecting to those countries experiencing turmoil. The baseline coefficient is reported first for comparison purposes. As we can see in Figures \ref{fig:conflict_leader} and \ref{fig:conflict_sanction}, the estimates when controlling for internal conflicts are virtually identical to our baseline estimates.

\clearpage
\subsection{Exclusion of one country at a time from the sample}

Below we show results when excluding one country at a time from the estimation. It is found that the coefficient hardly changes and that estimates are overall very stable. Moreover, the difference between our baseline coefficients and the simulated coefficients is never statistically significant (i.e. range of the values lie within the 95\% confidence intervals of the baseline coefficient).

\begin{figure}[ht!]
    \centering
    \begin{subfigure}[t]{0.45\textwidth}
        \centering
    	\includegraphics[width=\textwidth]{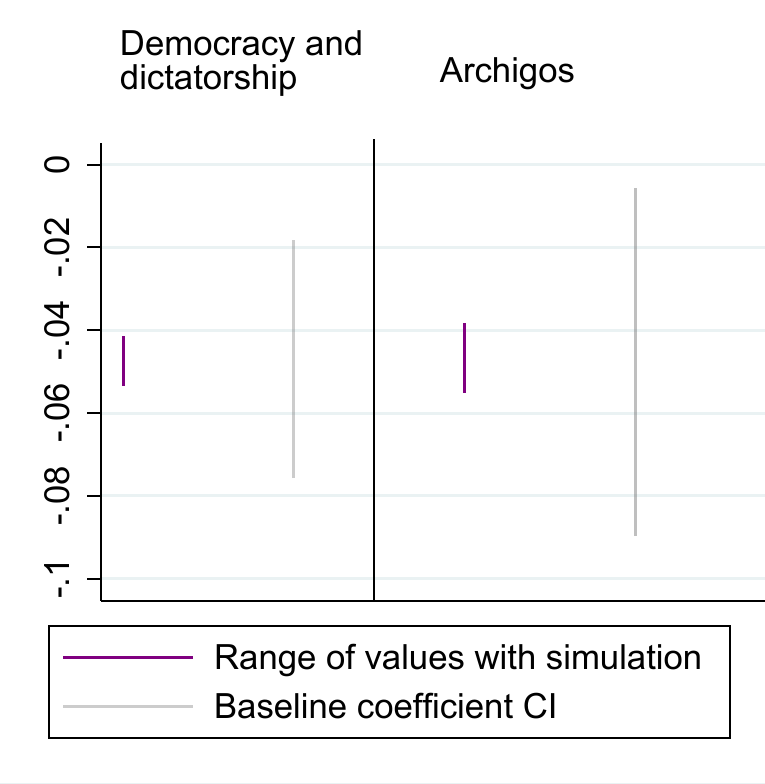}
        \caption{Leaders turnover}
    \end{subfigure}\\
    \begin{subfigure}[t]{0.6\textwidth}
        \centering
         \includegraphics[width=\textwidth]{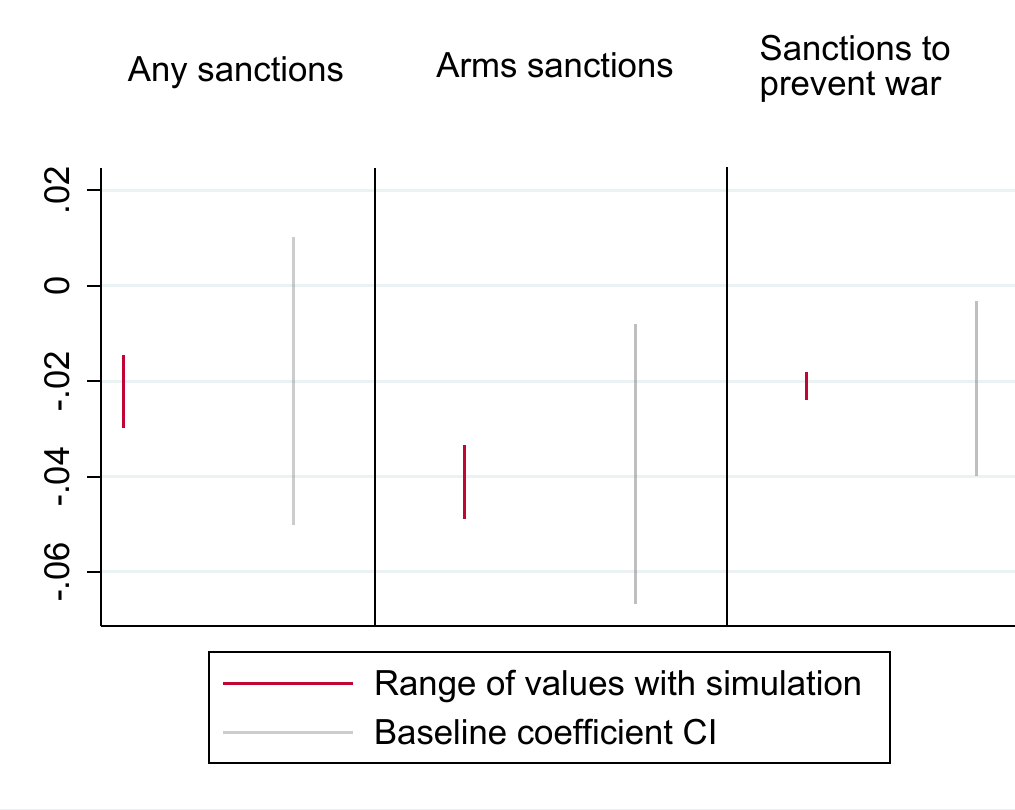}
        \caption{Sanctions }
    \end{subfigure}
	\caption{Comparison between the baseline model (with controls) coefficient confidence intervals and the range of values of the coefficients from the simulations (removing one country at a time)}
\footnotesize  \emph{Note: }Value and 95\% confidence intervals for the betweenness centrality coefficient. Linear Probability Models estimated with OLS. Standard errors clustered at the country level. The unit of observation is country-year and the dataset spans from 1993 to 2018 covering Africa, Asia and Europe.  The outcomes are, respectively, a dummy taking the value one if the executive leader changed during the year, and a dummy taking the value one if country $i$ was subject to sanctions in year $t$. Betweenness centrality corresponds to the natural logarithm of the natural gas pipeline betweenness centrality for country $i$. The vector of controls includes: degree centrality for country $i$, the log of the population, the log of the GDP per capita in ppp, the log of natural gas rents, as well as the log of oil rents both in percent of the GDP.
\end{figure}

\clearpage

\subsection{Exclusion of one pipeline at a time}

Below we show results when excluding one pipeline at a time from the estimation. It is found that the coefficient hardly changes and that estimates are overall very stable. Moreover, the difference between our baseline coefficients and the simulated coefficients is never statistically significant (i.e. the range of values lie within the 95\% confidence intervals of the baseline coefficient).

\begin{figure}[ht!]
    \centering
    \begin{subfigure}[t]{0.45\textwidth}
        \centering
    	\includegraphics[width=\textwidth]{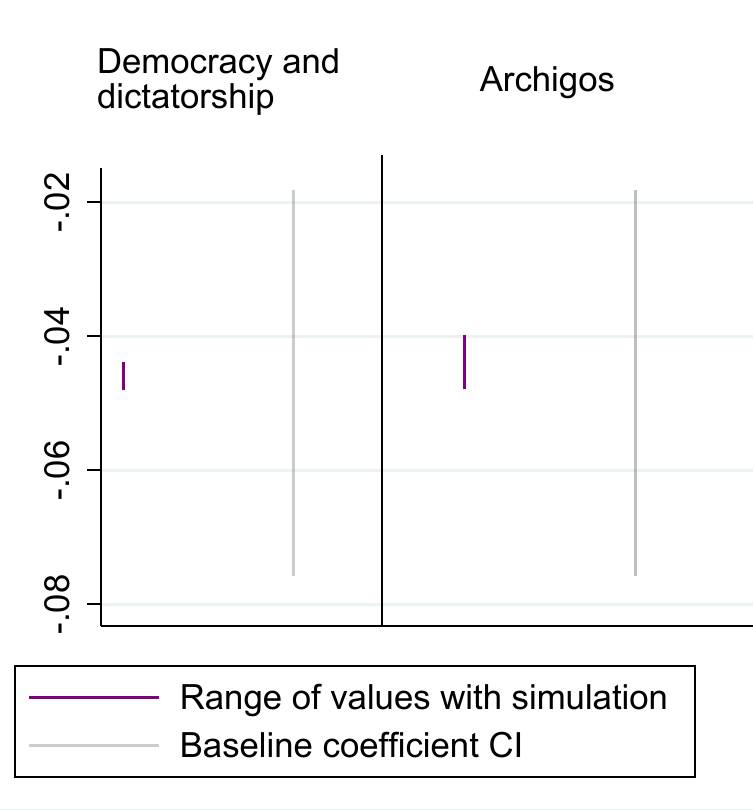}
        \caption{Leaders turnover}
    \end{subfigure}\\
    \begin{subfigure}[t]{0.6\textwidth}
        \centering
         \includegraphics[width=\textwidth]{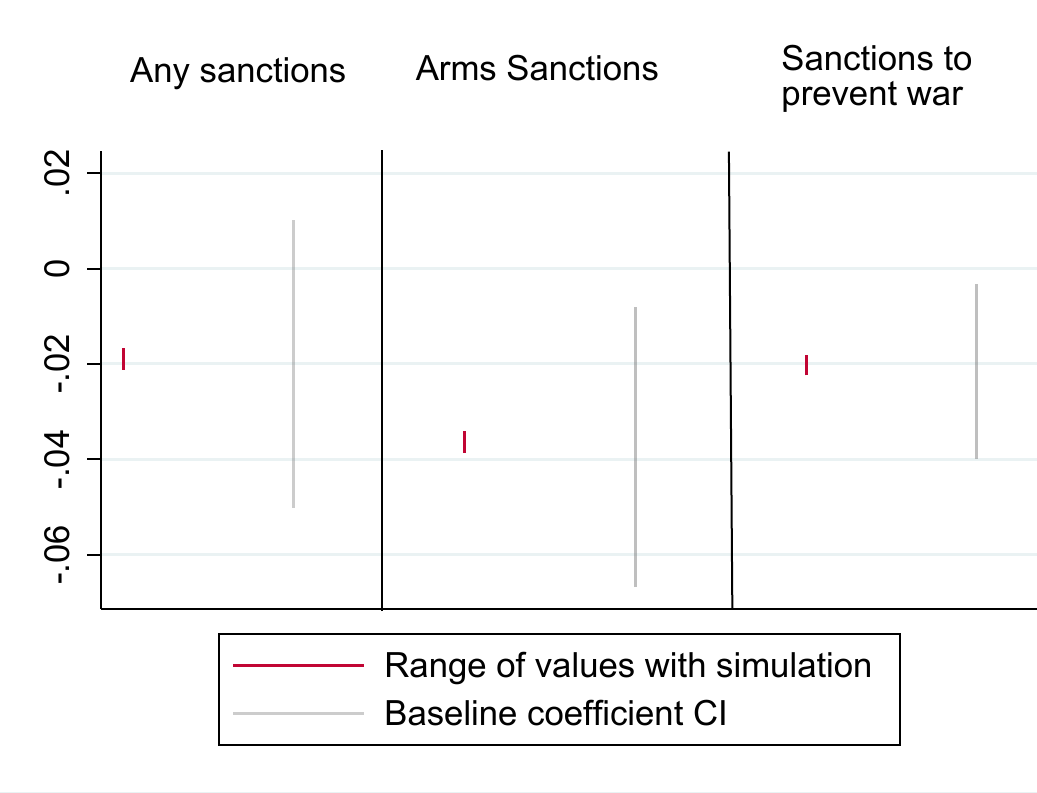}
        \caption{Sanctions }
    \end{subfigure}
	\caption{Comparison between the baseline model with controls coefficient confidence intervals and the range of values of the coefficient from the simulations (removing one country at a time)}
\footnotesize  \emph{Note: }Value and 95\% confidence intervals for the betweenness centrality coefficient. Linear Probability Models estimated with OLS. Standard errors clustered at the country level. The unit of observation is country-year and the dataset spans from 1993 to 2018 covering Africa, Asia and Europe.  The outcomes are, respectively, a dummy taking the value one if the executive leader changed during the year, and a dummy taking the value one if country $i$ was subject to sanctions in year $t$. Betweenness centrality corresponds to the natural logarithm of the natural gas pipeline betweenness centrality for country $i$. The vector of controls includes: degree centrality for country $i$, the log of the population, the log of the GDP per capita in ppp, the log of natural gas rents, as well as the log of oil rents both in percent of the GDP.
\end{figure}

\end{appendices}
 \end{document}